\documentclass[12pt]{article}
\usepackage{authblk}

\usepackage{hyperref}
\usepackage{fullpage}
\usepackage{datetime}
\usepackage{amsmath,amsfonts,amssymb, bm}

\usepackage{graphicx}

\newcommand{\pd}[2]{\frac{\partial #1}{\partial #2}}
\newcommand{\pdd}[2]{\frac{\partial^2 #1}{\partial  #2^2}}

\DeclareMathOperator{\diag}{diag}
\newcommand{\ee}{\mathbf{e}}
\newcommand{\p}{\mathbf{p}}
\newcommand{\q}{\mathbf{q}}
\newcommand{\g}{\mathbf{g}}
\newcommand{\x}{\mathbf{x}}

\newcommand{\vv}{\mathbf{v}}

\newcommand{\vvf}{\mathbf{v}_{\rm f}}
\newcommand{\vve}{\mathbf{v}_{\rm e}}

\newcommand{\vvs}{\mathbf{v}_{\rm s}}

\newcommand{\vs}{v_{\rm s}}
\newcommand{\us}{u_{\rm s}}
\newcommand{\ws}{w_{\rm s}}
\newcommand{\vf}{v_{\rm f}}
\newcommand{\uf}{u_{\rm f}}
 \newcommand{\wf}{w_{\rm f}}

\newcommand{\uu}{\mathbf{u}}
\newcommand{\uuf}{\mathbf{u}_{\rm f}}
\newcommand{\uue}{\mathbf{u}_{\rm e}}
\newcommand{\uus}{\mathbf{u}_{\rm s}}
\newcommand{\w}{\mathbf{w}}
\newcommand{\E}{\mathbf{E}}
\newcommand{\rr}{\mathbf{r}}
\newcommand{\n}{\hat{\mathbf{n}}}
\newcommand{\hn}{\hat{n}}
\newcommand{\nt}{\hat{\mathbf{n}}\tp}
\newcommand{\s}{\hat{\mathbf{s}}}
\newcommand{\ttt}{\hat{\mathbf{t}}}
\newcommand{\hatt}{\hat{\mathbf{t}}}
\newcommand{\hs}{\hat{s}}
\newcommand{\hht}{\hat{t}}
\newcommand{\e}{\hat{\mathbf{e}}}
\newcommand{\T}{\mathbf{T}}
\newcommand{\Tf}{\mathbf{T}_{\rm f}}
\newcommand{\boldS}{\mathbf{S}}
\newcommand{\I}{\mathbf{I}}
\newcommand{\boldd}{\mathbf{d}}
\newcommand{\boldtheta}{\bm{\theta}}

\newcommand{\sym}{\operatorname{sym}}

\newcommand{\sbb}[1]{[\![ #1 ]\!]} %
\newcommand{\ssb}[1]{[#1]} %

\renewcommand{\i}{\mathrm{i}}

\newcommand{\tp}{^{\mathsf{T}}} %
\DeclareMathOperator{\trace}{trace}

\newcommand{\R}{\mathbb{R}}

\renewcommand{\Re}{\operatorname{Re}}

\newcommand{\mufr}{\mu_{\mathrm{fr}}}
\newcommand{\mue}{\mu_{\mathrm{e}}}
\newcommand{\lambdae}{\lambda_{\mathrm{e}}}
\newcommand{\rhoe}{\rho_{\mathrm{e}}}

\newcommand{\kappafr}{\kappa_{\mathrm{fr}}}
\newcommand{\kappaf}{\kappa_{\mathrm{f}}}
\newcommand{\kappas}{\kappa_{\mathrm{s}}}
\newcommand{\rhoa}{\rho_{\mathrm{a}}}
\newcommand{\rhos}{\rho_{\mathrm{s}}}
\newcommand{\rhof}{\rho_{\mathrm{f}}}
\newcommand{\cpI}{c_p^{\mathrm{I}}}
\newcommand{\cpII}{c_p^{\mathrm{II}}}

\newcommand{\blockmatrix}[4]{
  \left(
    \begin{array}{c|c}
      #1 & #2 \\\hline
      #3 & #4 \\
    \end{array}
  \right)
}

\newcommand{\blockvector}[2]{
  \left(
    \begin{array}{c}
      #1 \\\hline
      #2
    \end{array}
  \right)
}

\begin{document}

 \title{A Discontinuous Galerkin method for three-dimensional
    elastic and poroelastic wave propagation: forward and adjoint problems}
    
\author[$\dagger$]{Nick Dudley Ward}
\author[$\star$]{Simon Eveson}
\author[$\ddag$]{Timo L\"ahivaara}

\affil[$\dagger$]{\footnotesize Department of Civil and Natural Resources Engineering, University of Canterbury, Christchurch, New Zealand, \texttt{nick.dudleyward@anu.edu.au}}
\affil[$\star$]{Department of Mathematics, University of York, York, United Kingdom, \texttt{simon.eveson@york.ac.uk}}
\affil[$\ddag$]{Department of Applied Physics, University of Eastern Finland, Kuopio, Finland, \texttt{timo.lahivaara@uef.fi}}
  
\maketitle

\begin{abstract}

We develop a numerical solver for three-dimensional wave propagation
in coupled poroelastic-elastic media, based on a high-order
discontinuous Galerkin (DG) method, with the Biot poroelastic wave
equation formulated as a first order conservative velocity/strain
hyperbolic system. To derive an upwind numerical flux, we find an
exact solution to the Riemann problem, including the
poroelastic-elastic interface; we also consider attenuation mechanisms
both in Biot's low- and high-frequency regimes. Using either a
low-storage explicit or implicit-explicit (IMEX) Runge-Kutta scheme,
according to the stiffness of the problem, we study the convergence
properties of the proposed DG scheme and verify its numerical
accuracy. In the Biot low frequency case, the wave can be highly
dissipative for small permeabilities; here, numerical errors
associated with the dissipation terms appear to dominate those arising
from discretisation of the main hyperbolic system.

We then implement the adjoint method for this formulation of Biot's
equation. In contrast with the usual second order formulation of the
Biot equation, we are not dealing with a self-adjoint system but, with
an appropriate inner product, the adjoint may be identified with a
non-conservative velocity/stress formulation of the Biot equation. We
derive dual fluxes for the adjoint and present a simple but
illuminating example of the application of the adjoint method.
\end{abstract}



\section{Introduction}

In \cite{dwle17} we solved the exact Riemann problem for coupled
poroelastic/elastic wave propagation in two dimensions and implemented
a solver in the discontinuous Galerkin (DG) framework developed in
\cite{hesthaven_warburton_book}. For the poroelastic case, we showed
that the usual convergence tests for an explicit time-marching scheme
were satisfied for a plane wave propagating through a square domain
provided the wave was not too dissipative (i.e. convergence order
$\sim$ order of polynomial basis plus 1 provided permeability is not
too small). In the case that the wave is too stiff (which corresponds
to a very small permeability and hence a very slow secondary P-wave)
the low storage Runge-Kutta scheme used in the explicit time-marching
scheme performed poorly, while a fourth order IMEX scheme developed in
\cite{Kennedy03} gave satisfactory results although proved sub-optimal
(i.e. convergence order $\sim$ order of polynomial basis minus 1).
We also showed that for a range of numerical examples our solver gave
accurate results and, in particular, resolved material
discontinuities.  In this paper we extend the method to
three-dimensional coupled poroelastic/elastic wave propagation.

Background information and references on numerical approaches to solving the
poroelastic wave equation are given in \cite{dwle17} and are not
repeated here. More recent work on numerical approaches to the poroelastic wave equation in the DF framework in three dimensions can be found in \cite{Shukla2020, Zhan18, Zhan2020}. We also provided background on our motivation for
studying poroelatic wave problems and the application to delineating
aquifers from ground motion data.

Apart from considering three-dimensional poroelastic wavefields the
current paper differs from our earlier paper \cite{dwle17} in one
major respect, since we develop the adjoint method for the poroelastic
wave equation using a first order formulation.  The adjoint method is
an extensively explored area, particularly in computational
seismology, since it is an approach to estimating derivatives of an
objective functional in a more economical fashion than simply running
multiple perturbations of the forward mapping, see for example
\cite{tromp08} and \cite{fichtner11}. A second order formulation of a
wave equation is self-adjoint and therfore presents little
difficulty. For a first order formulation this is no longer the case
and more care has to be taken to obtain the adjoint wavefield as well
as numerical fluxes.  For the elastic and other simpler wave equations
this has been considered in \cite{wilcox15}. In this paper we consider
the adjoint method for coupled elastic/poroelatic problems and derive
appropriate fluxes.

The structure of this paper is as follows. First, in Section
\ref{sec:biot} we present a formulation of Biot's equations. In
Section \ref{sec:numer-scheme-invisc} we describe the DG scheme used
in this study including a derivation of upwind fluxes based on a
solution of the associated Riemann problem. In Section
\ref{sec:dissip} we consider poroelasticity, and in Section
\ref{sec:elast-coupl} we derive upwind fluxes for coupled
elastic/poroelastic models. Next in Section \ref{sec:adjoint} we
discuss the adjoint method for the first order hyperbolic formulation
of the poroelastic wave equation and derive dual upwind numerical
fluxes for the adjoint poroelastic wavefield. In section
\ref{sec:numer-exper} we present numerical experiments including a
convergence study. Finally a discussion and concluding remarks are
given in Sections \ref{sec:discussion} and \ref{sec:conclusions}
respectively.

\section{Biot's equations of motion for poroelastic wave propagation}
\label{sec:biot}

In this section we formulate Biot's equations of motion for
poroelastic wave propagation given in the classical papers
\cite{biot56a} and \cite{biot56b}. A more detailed account can be
found in \cite{dwle17} or \cite{carcione15}.

Denote by $\uus$ the solid displacement, by $\uuf$ the fluid
displacement, and by $\w$ the relative displacement of fluid
$\w=\phi(\uuf -\uus )$, where $\phi$ is porosity.  Note that $\w$ is
volumetric flow per unit area of the bulk medium. Then Biot's
equations of poroelastic wave propagation for the laminar case may be
stated as

\begin{align}
  \label{eq:biot2a}
  \rhoa \pdd{\uus}{t} +\rhof \pdd{\w}{t} &=\nabla\cdot {\bf T},\\
  \rhof \pdd{\uus}{t} + m \pdd{\w}{t} + \frac{\eta}{k}\pd{\w}{t}
  &=\nabla\cdot \Tf, \label{eq:biot2b}
\end{align}
where $\rhos$ is the solid density, $\rhof$ the fluid density, $\rhoa$ is the average density
$$
\rhoa = (1-\phi)\rhos + \phi \rhof
$$
and
\begin{equation}
  m = \rhof \tau / \phi
\end{equation}
where $\tau$ is the fluid tortuosity and $\phi$ the porosity. The
coefficient of the dissipative term $\displaystyle{\pd{\w}{t}}$ is the
ratio of the viscosity $\eta$ to the permeability $k$ of the porous
medium.  The stress tensors ${\bf T}$
and $\Tf$ are isotropic Hooke's laws and are discussed in the next
section.  For a detailed derivation see \cite{carcione15}.

The most distinctive feature of Biot's early papers \cite{biot56a,
  biot56b} is the existence of a characteristic frequency $f_c$, below
which the Pouiselle assumption is valid and inertial forces are
negligible to viscous forces:
\begin{equation}
  f_c = \frac{\eta \phi}{2 \pi \tau \rhof k}.
\end{equation}
See \cite{carcione15}, Section 7.6.1.  At higher frequencies, inertial
forces are no longer negligible, and the viscous resistance to fluid
flow given by the coefficient of the dissipative term is
frequency-dependent. In \cite{biot62a} Biot introduced a viscodynamic
operator to model the high frequency regime.

\subsection{Poroelastic Hooke's laws}

In \cite{biot56a} Biot proposed generalised Hooke's laws to describe
the stress-strain coupling between solid and fluid.  Letting $\mathbf{E}$
denote the solid strain tensor
\begin{equation}
\label{eq:strain_tensor}
\E = \frac{1}{2}(\nabla \uus +(\nabla\uus)\tp)
\end{equation}
and $\epsilon = \nabla\cdot\uuf$ the strain in the fluid, these may be stated in the form:
\begin{eqnarray}
  (1-\phi) {\bf T}_{\rm s} &=& 2\mu {\bf E} + \lambda \trace({\bf E}){\bf I} + Q \epsilon {\bf I}\\
  \phi \Tf &=& Q \trace({\bf E}){\bf I} + M \epsilon {\bf I}
\end{eqnarray}
where $\mu$ and $\lambda$ correspond to the usual Lam{\' e}
coefficients, and ${\bf I}$ denotes the identity tensor.  As usual,
under the assumption that the fluid does not support shear stress, one
may interpret $\mu$ as the dry matrix shear modulus $\mufr$.

Biot and Willis \cite{biot_willis57} showed that the elasticity
coefficients postulated above may be written in terms of bulk moduli
defined by idealised experiments, viz. the frame bulk modulus of the
frame $\kappafr$, the bulk modulus of the solid $\kappas$ and the bulk
modulus of the fluid $\kappaf$.  Carcione gives a detailed account in
\cite{carcione15}. Since we are interested in the system
(\ref{eq:biot2a})--(\ref{eq:biot2b}), we may write
\begin{eqnarray}
  {\bf T} &=& 2\mufr {\bf E} + \left(B-\frac{2}{3}\mufr\right)\trace({\bf E}){\bf I} - C\zeta {\bf I} \label{hooke1}\\
  \Tf &=& C\trace({\bf E}){\bf I} - M\zeta {\bf I} \label{solid_tensor}
\end{eqnarray}
where ${\bf T} = (1-\phi){\bf T}_s + \phi\Tf$ is total stress and
$\zeta = -\nabla\cdot\w$ is the variation of fluid content. The moduli
$B,C,$ and $M$ can be written as
\begin{align}
  B &=\frac{\kappas - (1+\phi)\kappafr + \phi\kappas\kappafr/\kappaf}
  {(1-\kappafr/\kappas) - \phi(1-\kappas/\kappaf)}, \label{bigB}\\
  C &=\frac{(1-\kappafr/\kappas)\kappas}
  {(1-\kappafr/\kappas) - \phi(1-\kappas/\kappaf)},
\end{align}
and
\begin{equation}
  M =\frac{\kappas}
  {(1-\kappafr/\kappas) - \phi(1-\kappas/\kappaf)}. \label{bigM}
\end{equation}

One of the less desirable aspects of poroelastic theory is the
proliferation of constants.  A neater formulation that is possibly
better suited to estimation is to introduce the Biot effective stress
constant $\alpha$ given by

$$
\alpha = 1 - \frac{\kappafr}{\kappas}.
$$
Then we can write the solid and fluid stress tensors as
\begin{eqnarray}
  {\bf T} &=& 2\mufr {\bf E} + \left(\kappafr + \alpha^2 M -\frac{2}{3}\mufr\right)\trace({\bf E}){\bf I} 
  - \alpha M\zeta {\bf I}\\
  \Tf &=& M(\alpha\trace({\bf E}) - \zeta) {\bf I}.
\end{eqnarray}

\section{Numerical scheme for the inviscid case}\label{sec:numer-scheme-invisc}

\subsection{Hyperbolic system}
\label{sec:hypsys}

We use a velocity-strain formulation to express
(\ref{eq:biot2a})--(\ref{eq:biot2b}) as a first-order conservative
hyperbolic system.  Introducing the variable

\begin{equation}
  \label{hypsys_cf}
  \q = (\epsilon_{11}, \epsilon_{22},\epsilon_{33}, \epsilon_{12}, \epsilon_{23}, \epsilon_{13},\zeta, \us, \vs, \ws, \uf, \vf,\wf)\tp
\end{equation}
where the $\epsilon_{ij}$ are components of the solid strain tensor,
$\zeta$ is the variation of fluid content, $\vvs=(\us,\vs,\ws)$ are the $x$, $y$ and
$z$ components of the solid velocity $\displaystyle{\pd{\uus}{t}}$ and
$\vvf = (\uf,\vf,\wf)$ are the components of the relative fluid velocity
$\displaystyle{\pd{\w}{t}}$, viz.
\begin{equation}
  \E =
  \begin{pmatrix}
    \epsilon_{11} &\epsilon_{12} & \epsilon_{13}\\
    \epsilon_{12} &\epsilon_{22} & \epsilon_{23}\\
    \epsilon_{13} &\epsilon_{23} & \epsilon_{33}\\
  \end{pmatrix}
\end{equation}
and
\begin{align}
  \zeta &= -\nabla\cdot\w\\
  (\us,\vs,\ws)\tp & =\pd{\uus}{t}\\
  (\uf,\vf,\wf)\tp & =\pd{\w}{t}\\
\end{align}
we obtain, using the Einstein summation convention
\begin{equation}
  \label{hyp_sys_mat}
  Q \pd{\q}{t} + \nabla\cdot\mathcal{F} 
  = Q \pd{\q}{t} + \pd{(A^i \q)}{x_i} = \g+\g_V %
\end{equation}
Here $\mathcal{F}$, $Q$, $A^i$, $\g$ and $\g_V$ are as follows:
$$
\mathcal{F} = [F_1, F_2, F_3] = [A^1 \q, A^2 \q, A^3 \q]
$$

\begin{equation}
  \label{eq:mass_matrix}
  Q =
\blockmatrix{Q_1}{0}{0}{Q_2}
\end{equation}
where $Q_1$ is the $7\times 7$ identity matrix and
\begin{equation}
  Q_2 =
  \begin{pmatrix}
    
    \rhoa &0     &0  &\rhof & 0     &0\\
    0     &\rhoa &0  &0 & \rhof &0\\
    0     &0     &\rhoa  &0 & 0 &\rhof\\
    \rhof &0     &0 &m &0 &0\\ 
    0     &\rhof &0 &0 &m &0\\
    0     &0     &\rhof &0 &0 &m\\
  \end{pmatrix}.
\end{equation}
The Jacobian matrices $A^i$, $i=1,2,3,$ may similarly be given in block form
\begin{equation}
  \label{eq:jac_mat}
  A^i = \blockmatrix{0}{A_2^i}{A_1^i}{0}
\end{equation}
where the matrices $A^i_1$ and $A^i_2$ are in Table~\ref{tab:Jacobian-block}.
\begin{table}
\caption{The off-diagonal blocks of the Jacobian matrices $A^i$. Here
$\lambda = \kappafr + \alpha^2 M -\frac{2}{3}\mufr$.}
\label{tab:Jacobian-block}
\begin{align*}
  A_1^1 & = - \begin{pmatrix}
    2\mufr + \lambda & \lambda &  \lambda & 0 & 0 &0 & -\alpha M \\
     0               & 0       &  0 & 2\mufr & 0 &0 & 0 \\
     0               & 0       &  0 & 0      & 0 &2\mufr & 0\\
     \alpha M& \alpha M& \alpha M& 0& 0& 0&  -M\\
     0      & 0 & 0 & 0 &  0 &0 &0 \\
    0 & 0 & 0 & 0 &  0 &0 &0\\
                \end{pmatrix}
        & A_2^1 = - \begin{pmatrix}
           1 & 0 & 0 & 0 & 0 & 0 \\
           0 & 0 & 0 & 0 & 0 & 0 \\
           0 & 0 & 0 & 0 & 0 & 0 \\ 
           0 & 1/2 & 0 & 0 & 0 & 0 \\ 
           0 & 0 & 0 & 0 & 0 & 0 \\ 
           0 & 0 & 1/2 & 0 & 0 & 0 \\ 
           0 & 0 & 0   & -1 & 0 & 0 \\
                    \end{pmatrix}\\
  A_1^2 &= - \begin{pmatrix}
    0            & 0       &  0 & 2\mufr & 0 &0 & 0\\
 \lambda         &2\mufr + \lambda & \lambda & 0 & 0 &0 & -\alpha M\\
     0               & 0       &  0 & 0      & 2\mufr &0 & 0\\
     0 & 0 & 0 & 0 &  0 &0 &0\\
    \alpha M& \alpha M& \alpha M& 0& 0& 0&  -M &\\
    0 & 0 & 0 & 0 &  0 &0 &0\\
  \end{pmatrix}
  & A_2^2 = - \begin{pmatrix}
    0 & 0 & 0 & 0 & 0 & 0 \\
    0 & 1 & 0 & 0 & 0 & 0 \\
    0 & 0 & 0 & 0 & 0 & 0 \\ 
    1/2 & 0 & 0 & 0 & 0 & 0 \\ 
    0 & 0 & 1/2 & 0 & 0 & 0 \\ 
    0 & 0 & 0   & 0 & 0 & 0 \\ 
    0 & 0 & 0   & 0 & -1 & 0 \\ 
               \end{pmatrix}\\
  A_1^3 &= - \begin{pmatrix} 
    0 & 0 & 0 & 0 & 0 &2\mufr & 0\\
    0 & 0 & 0 & 0 & 2\mufr &0      & 0\\
     \lambda         & \lambda &  2\mufr + \lambda & 0      & 0 &0  & -\alpha M\\
    0 & 0 & 0 & 0 &  0 &0 &0\\
    0 & 0 & 0 & 0 &  0 &0 &0\\
    \alpha M& \alpha M& \alpha M & 0& 0& 0&  -M\\
  \end{pmatrix}
  & A_2^3 = - \begin{pmatrix}
    0 & 0 & 0 & 0 & 0 & 0 \\
    0 & 0 & 0 & 0 & 0 & 0 \\
    0 & 0 & 1 & 0 & 0 & 0 \\ 
    0 & 0 & 0 & 0 & 0 & 0 \\ 
    0 & 1/2 & 0 & 0 & 0 & 0 \\ 
    1/2 & 0 & 0 & 0 & 0 & 0 \\ 
    0 & 0 & 0   & 0 & 0 & -1 \\ 
  \end{pmatrix}
\end{align*}
\end{table}
For the low-frequency dissipative regime considered in Section
\ref{sec:dissip} the source term $\g$ is given by
\begin{equation}
  \label{low_freq_source}
   \mathbf{g} = (\mathbf{0}_{10},-\frac{\eta}{k}\uf, -\frac{\eta}{k}\vf,  -\frac{\eta}{k}\wf)\tp
 \end{equation}
 where $\mathbf{0}_{10}$ is a $1 \times 10$ zero row vector and $\g_V$ is a
 volume source defined in Section \ref{sec:numer-exper}.

The eigenstructure of $Q^{-1} A^1$ is derived in detail in the
appendix of \cite{dwle17} and summarised below. Introducing the
quantities
\begin{align}
  Z_1 &= m\rhoa -\rhof^2\\
  Z_2 &= -2\rhof\alpha M + \rhoa M + m\lambda + 2m\mufr\\
  Z_3 &= \rhoa(4\alpha^2 m-4\alpha\rhof +\rhoa)M^2 -2(2\alpha m\rhof +m\rhoa -2\rhof^2)M(2\mufr+\lambda) + m^2(2\mufr+\lambda)^2\\
  Z_4 &= \rhoa M - m\lambda - 2m\mufr\\
  Z_5 & = 2(\alpha m - \rhof)M
\end{align} 
we have the following expressions for the wave speeds for the
non-dissipative case:
\begin{align}
  \cpI    &= \pm\sqrt{\frac{Z_2 + \sqrt{Z_3}}{2Z_1}}\\
  \cpII &= \pm\sqrt{\frac{Z_2 - \sqrt{Z_3}}{2Z_1}}\\
  c_s       &= \pm\sqrt{\frac{m\mufr}{Z_1} }.
\end{align}
Here $\cpI$ is the speed of the fast P-wave corresponding to the
P-wave of ordinary elasticity, $\cpII$ is Biot's slow P-wave, and
$c_s$ is the speed of the shear wave, where usually
$\cpI > c_s >\cpII$.  Writing
$\displaystyle{\Lambda = \diag(-\cpI, -c_s, -c_s, -\cpII, \cpII, c_s,
  c_s, \cpI)}$ for the non-zero eigenvalues of $Q^{-1}A^1$
corresponding representative eigenvectors are given by the columns of
\begin{equation}
  \label{ev_matrix}
  R =
  \begin{pmatrix}
    1  &0  &0 &1 &1 &0 &0 &1 \\
    0  &0  &0 &0 &0 &0 &0 &0 \\
    0  &0  &0 &0 &0 &0 &0 &0 \\
    0  &1/2 &0 &0 &0 &0 &1/2 &0 \\
    0  &0  &0 &0 &0 &0 &0 &0 \\
    0  &0  &1/2 &0 &0 &1/2 &0 &0 \\
    -\gamma_1  &0  &0 &-\gamma_2 &-\gamma_2 &0 &0 &-\gamma_1 \\
    \cpI  &0  &0 &\cpII &-\cpII &0 &0 &-\cpI \\
    0  &c_s  &0 &0 &0 &0 &-c_s &0 \\
    0  &0  &c_s &0 &0 &-c_s &0 &0 \\
    \gamma_1\cpI  &0  &0 &\gamma_2\cpII &-\gamma_2\cpII &0 &0 &-\gamma_1\cpI \\
    0  &-c_s \rho_f/m  &0 &0 &0 &0 &c_s\rho_f/m &0 \\
    0  &0  &-c_s \rho_f/m &0 &0 &c_s \rho_f/m &0 &0 \\
  \end{pmatrix}
\end{equation}
where $\gamma_1 = (Z_4 + \sqrt{Z_3})/Z_5$ and
$\gamma_2 = (Z_4 - \sqrt{Z_3})/Z_5$. 

\subsection{Discontinuous Galerkin method}

In this section we outline the DG method.  Our formulation follows
Hesthaven and Warburton \cite{hesthaven_warburton_book}, where a
detailed account of the DG method can be found.  We first suppose that
the computational domain $\Omega\subset \mathbb{R}^3$ is devided into tetrahedra 
using $K$ elements
$$
\Omega = \bigcup_{k=1}^K D^k.
$$
The boundary of element $D^k$ is denoted by $\partial D^k$.  We assume
that the elements are aligned with material discontinuities.
Furthermore, for any element $D^k$ the superscript `$-$' refers to
interior information while `$+$' refers to exterior information.

To obtain the strong form we multiply (\ref{hyp_sys_mat}) by a local
test function $p^k$ and integrate by parts twice to obtain an
elementwise variational formulation
  
\begin{equation}
\label{strong_form}
  \int_{D^k}  \left(Q \pd{\q^k}{t} 
    + \nabla\cdot\mathcal{F} -\g -\g_V \right) p^k dx 
  =  \oint_{\partial D^k} 
  \n\cdot(\mathcal{F}^- - \mathcal{F}^*) p^k d\Gamma,
\end{equation}
where $\n$ is an outward pointing unit normal, $\q^k$ is the
restriction of $\q$ to the element $D^k$ and $\mathcal{F}^*$ is the
numerical flux across neighbouring element interfaces.  To discretise
(\ref{strong_form}) the elementwise solutions $\q^k$ and the test
functions $p^k$ are approximated using the same polynomial basis
functions \cite{hesthaven_warburton_book}.

To approximate the numerical flux $\mathcal{F}^*$ along the normal
$\n$ we solve the Riemann problem at an interface. With this in mind we
define
$$
\Pi = \hat{n}_x A^1 + \hat{n}_y A^2 +\hat{n}_z A^3
$$
so that
$$
\n\cdot\mathcal{F} = \Pi \q
$$

\subsection{Boundary conditions}\label{sec:boundary-conditions}

The ground surface of the porous medium is modelled as as
traction-free surface, viz.  ${\bf T}\n = \Tf\n = \mathbf{0}$ while
other boundaries are modelled as Dirichlet or absorbing boundaries.
The latter are implemented as outflows by setting the flux equal to
zero.  This is only exact for one-dimensional problems and may
introduce boundary artefacts.
 
\subsection{Riemann problem}\label{sec:riem}

Now that the eigenstructure of $Q^{-1}A^1$ has been established we
proceed to solve the Riemann problem for (\ref{hyp_sys_mat}) using
the same calculations carried out in \cite{dwle17}.

In the following calculations it is convenient to work with a local
interface basis $\{\n, \s, \hatt\}$ where $\s$, $\hatt$ are orthogonal
unit tangent vectors.  Using a prime to denote vectors with respect to
the interface basis, we write $\q = L\q^\prime$ where $L$ is the
change of basis map from $\{\n,\s, \hatt \}$ to the physical Euclidean
basis $\{\e_1,\e_2,\e_3\}$.  It is straightforward to show that
\begin{equation}
  \label{local_coord}
  \q^\prime = L^{-1}\q =(\n\tp\E\n, \s\tp \E\s, \hatt\tp \E\hatt, \s\tp\E\n, \hatt\tp\E\s, \hatt\tp\E\n, \zeta,
  \n\cdot \vvs, \s\cdot\vvs, \hatt\cdot\vvs,\n\cdot\vvf, \s\cdot\vvf,
  \hatt\cdot\vvf)\tp.
\end{equation}
Letting $P=[\n\,\, \s \,\, \hatt]$ the first three
terms follow from the change of basis formula for a matrix
$\E^\prime=P\tp \E P$, and the last four terms follow from $\vv^\prime
= P\tp\vv$.

We also have
\begin{equation}
  \label{jac_mat_trans}
  L^{-1}\Pi L = A^1  \text{ and } L^{-1}Q^{-1}\Pi L = Q^{-1}A^1
\end{equation}

\begin{figure}[!h]
  \centering
  \includegraphics[width=0.49\textwidth]{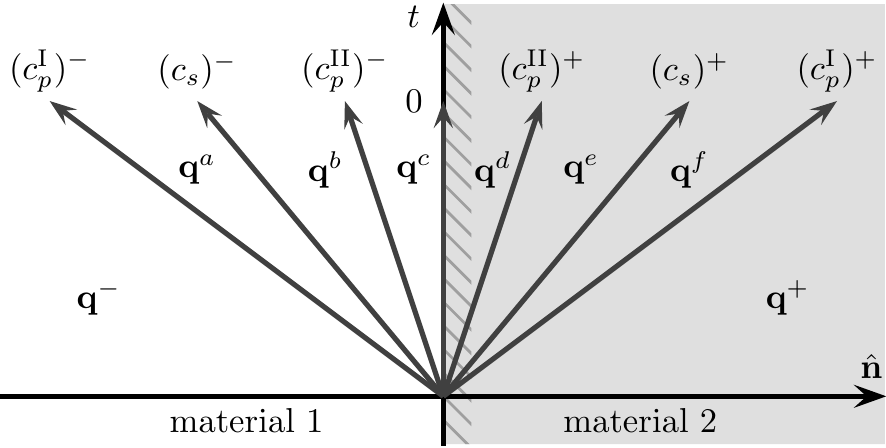}

  \caption{Schematic showing characteristic wave speeds at a poroelastic
    interface between two states $\q^-$ and $\q^+$. $\q^a$-- $\q^f$
    denote the intermediate states.}
  \label{fig:riemann_prob}
\end{figure}

To compute an upwind numerical flux across an interface for the
two-dimensional locally isotropic poroelastic system (\ref{hypsys_cf})
we solve a Riemann problem at an interface.  This consists of solving
the system (\ref{hypsys_cf}) with initial data
$$
\q_0(\x) =
\begin{cases}
  \q^- & \text{if }  \n\cdot (\x-\x_0)<0\\
  \q^+ & \text{if } \n\cdot (\x-\x_0)>0
\end{cases}
$$
where $\x_0$ is a point on the interface.  
For each wave speed $c$, the Rankine-Hugoniot jump condition,
\cite{hesthaven_warburton_book,leveque}
$$
-c Q[\q^- - \q^+] + [ (\Pi \q)^- - (\Pi \q)^+] = 0
$$
holds across each wave, where the superscripts $-$ and $+$ refer
respectively to the interior and exterior information on an element.
We have six unknown states $(\q^a, \q^b, \q^c, \q^d, \q^e, \q^f)$
shown in Figure~\ref{fig:riemann_prob}, with the following jump
conditions:
\begin{align}
  (\cpI)^- Q^- (\q^- - \q^a) + \Pi^- (\q^- - \q^a)  &=0\\
  (c_s)^-    Q^- (\q^a - \q^b) + \Pi^- (\q^a - \q^b) &=0\\
  (\cpII)^- Q^- (\q^b - \q^c) + \Pi^- (\q^b - \q^c) &=0\\
  \Pi^- \q^c - \Pi^+ \q^d &=0 \label{couple_interface}\\
  -(\cpII)^+ Q^+(\q^d - \q^e) + \Pi^+ (\q^d - \q^e)  &=0\\
  -(c_s)^+    Q^+ (\q^e - \q^f) + \Pi^+ (\q^e - \q^f) &=0\\
  -(\cpI)^+  Q^+ (\q^f - \q^+) + \Pi^+ (\q^f- \q^+) &=0
\end{align}

Thus:

\begin{align}
  \q^- - \q^a &= \beta_1 \rr_1^- \\
  \q^a - \q^b &= \beta_2 \rr_2^- + \beta_3 \rr_3^-\\
  \q^b - \q^c &= \beta_4 \rr_4^- \\
  \q^d - \q^e &= \beta_{10} \rr_{10}^+ \\
  \q^e - \q^f &= \beta_{11} \rr_{11}^+ + \beta_{12} \rr_{12}^+ \\
  \q^f - \q^-&= \beta_{13} \rr_{13}^+ 
\end{align}
where $\rr_j^\pm$ is an eigenvector corresponding to wave speed
$c_j^\pm$ and hence
\begin{align}
  \q^- - \q^c &= \beta_1 \rr_1^- + \beta_2 \rr_2^- +\beta_3 \rr_3^- + \beta_4 \rr_4^+ \label{link_1}\\
  \q^d - \q^+ &= \beta_{10} \rr_{10}^+ + \beta_{11} \rr_{11}^+ +\beta_{12} \rr_{12}^+ +\beta_{13} \rr_{13}^+ \label{link_2}
\end{align}
Note that $\rr_4, \ldots, \rr_9$ correspond to wavespeed zero and are not referenced in the following derivations.

We now make use of the orthogonality of the P-wave and the S-wave
eigenvectors to uncouple the system (\ref{link_1}) and (\ref{link_2}).
Recall that the eigenvectors $\rr_1^-, \rr_{13}^+$ correspond to fast
P-waves, $\rr_4^-, \rr_{10}^+$ to slow P-waves, and
$\rr_2^-, \rr_3^+$, $\rr_{11}^+$, $\rr_{12}^+$ to S-waves.  First we
deal with the P-wave coefficients
$\beta_1, \beta_3, \beta_{10}, \beta_{13}$.

From the interface condition (\ref{couple_interface}) we have
$$
\Pi^- \q^c = \Pi^+ \q^d
$$
and so
$$
L^{-1}\Pi^-\q^c = L^{-1}\Pi^+ \q^d.
$$
Using the first equality in (\ref{jac_mat_trans}) this gives
$$
A^- (L^{-1} \q^c) = A^+ (L^{-1} \q^d),
$$
that is
\begin{equation}
  \label{cont_flux}
  A^- (\q^{c})^\prime = A^+ (\q^{d})^\prime.
\end{equation}
Recalling that
$$
{\bf T}^\pm = 2\mufr^\pm {\bf E} + \lambda^\pm\trace({\bf E}){\bf I} - \alpha^\pm M^\pm\zeta {\bf I}
$$
where $\displaystyle{\lambda^\pm =\kappafr^\pm + \alpha^{2\pm}
  M^{\pm} -\frac{2}{3}\mufr^{\pm}}$ and the $\pm$ indicates
whether $\T$ is evaluated on the interior or exterior of the
interface, it follows that

\begin{align}
  \n\tp \T^\pm \n &= 2\mufr^\pm \n\tp\E^\pm\n + \lambda^\pm \trace(\E^\pm)\n\tp \I\n -\alpha^\pm M^\pm\zeta^\pm \n\tp\I\n \nonumber\\
  &= 2\mufr^\pm \n\tp\E^\pm\n + \lambda^\pm \trace(\E^\pm) -\alpha^\pm M^\pm\zeta^\pm \nonumber\\
  &=  2\mufr^\pm \n\tp\E^\pm\n + \lambda^\pm (\n\tp\E^\pm\n + \s\tp\E^\pm\s + \hatt\tp\E^\pm\hatt) -\alpha^\pm M^\pm\zeta^\pm \label{iden_1}
\end{align}
since the trace is invariant under orthogonal transformations.  We
also have

\begin{equation}
  \s\tp \T^\pm \n = 2\mufr\s\tp \E^\pm\n,\quad \hatt\tp \T^\pm \s = 2\mufr\hatt\tp \E^\pm\s, \quad \hatt\tp \T^\pm \n = 2\mufr\hatt\tp \E^\pm\n.
  \label{iden_1b}
\end{equation}

We obtain similarly for
$$
\Tf^\pm = M^\pm(\alpha^\pm\trace({\bf E}^\pm) - \zeta^\pm) {\bf I}
$$
the following identity:

\begin{align}
  \n\tp \T_{\rm f}^\pm \n &=  M^\pm\alpha^\pm\trace(\E^\pm) - M^\pm\zeta^\pm \nonumber\\
  &= M^\pm\alpha^\pm(\n\tp\E^\pm\n + \s\tp\E^\pm\s + \hatt\tp\E^\pm\hatt) - M^\pm\zeta^\pm.
  \label{iden_2}
\end{align}

Also

\begin{equation}
  \s\tp \T^\pm_{\rm f} \n = \hatt\tp \T^\pm_{\rm f} \s = \hatt\tp \T^\pm_{\rm f} \n = 0.
\end{equation}

From (\ref{cont_flux}) we obtain the following flux continuity
relations
\begin{align}
  \n\cdot\vvs^c &= \n\cdot\vvs^d\label{cont_cond_a}\\
  \s\cdot\vvs^c &= \s\cdot\vvs^d\label{cont_cond_b}\\
  \hatt\cdot\vvs^c &= \hatt\cdot\vvs^d\\
  \n\cdot\vvf^c &= \n\cdot\vvf^d \label{cont_cond_c}\\
  \n\tp \T^c \n &= \n\tp \T^d \n \label{cont_cond_d}\\
  \s\tp\T^c\n &= \s\tp\T^d\n\label{cont_cond_e}\\
  \hatt\tp\T^c\n &= \hatt\tp\T^d\n\\
  \n\tp \T_{\rm f}^c \n &= \n\tp \T_{\rm f}^d \n \label{cont_cond_f}
\end{align}
where we have used (\ref{iden_1}), (\ref{iden_1b}) and (\ref{iden_2}).

We now proceed with the evaluation of the $\beta$ terms. From
(\ref{link_1}) we have

$$
L^{-1} \q^- - L^{-1} \q^c = \beta_1 (\rr_1^\prime)^- + \beta_2 (\rr_2^\prime)^- + \beta_3 (\rr_3^\prime)^- + \beta_4 (\rr_4^\prime)^- 
$$
where the $(\rr_j^\prime)^-$ are the $j$'th columns of the eigenvector
matrix $R$ given by equation (\ref{ev_matrix}) evaluated in the
interior of an element.  Unwrapping, and using (\ref{local_coord}), we
obtain the relationships

\begin{align}
  \nt \E^- \n - \nt \E^c\n &= \beta_1 + \beta_4 \label{lhs_a}\\
  \s\tp\E^-\s &= \s\tp\E^c\s \label{lhs_b}\\
  \hatt\tp\E^-\hatt &= \hatt\tp\E^c\hatt\\
  \s\tp\E^-\n -\s\tp\E^c\n &= \beta_2/2 \label{lhs_c}\\
  \hatt\tp\E^-\s &= \hatt\tp\E^c\s\\
  \hatt\tp\E^-\n -\hatt\tp\E^c\n &= \beta_3/2\\
  \zeta^- - \zeta^c &= -\gamma_1^- \beta_1  -\gamma_2^-\beta_4 \label{lhs_d}\\
  \n\cdot\vvs^- - \n\cdot\vvs^c &= (\cpI)^-\beta_1 + (\cpII)^-\beta_4 \label{lhs_e}\\
  \s\cdot\vvs^- - \s\cdot\vvs^c &= (c_s)^-\beta_2 \label{lhs_f}\\
  \hatt\cdot\vvs^- - \hatt\cdot\vvs^c &= (c_s)^-\beta_3 \\
  \n\cdot\vvf^- - \n\cdot\vvf^c &= (\gamma_1 \cpI)^-\beta_1 + (\gamma_2 \cpII)^-\beta_4 \label{lhs_g}\\
  \s\cdot\vvf^- - \s\cdot\vvf^c &= -(c_s\rhof/m)^-\beta_2\\
  \hatt\cdot\vvf^- - \hatt\cdot\vvf^c &= -(c_s\rhof/m)^-\beta_3
\end{align}
We derive similar relations on the right-hand side. From
(\ref{link_2}) we have
$$
L^{-1} \q^d - L^{-1} \q^+ = \beta_{10} (\rr_{10}^\prime)^+ + \beta_{11} (\rr_{11}^\prime)^+ + \beta_{12}
(\rr_{12}^\prime)^+ + \beta_{13} (\rr_{13}^\prime)^+
$$
Thus:

\begin{align}
  \nt \E^d \n - \nt \E^+\n &= \beta_5 + \beta_8 \label{rhs_a}\\
  \s\tp\E^d\s &= \s\tp\E^+\s \label{rhs_b}\\
  \hatt\tp\E^d\hatt &= \hatt\tp\E^+\hatt \\
  \s\tp\E^d\n -\s\tp\E^+\n &= \beta_{12}/2 \label{rhs_c}\\
  \hatt\tp\E^d\s &= \hatt\tp\E^+\s \\
  \hatt\tp\E^d\n -\hatt\tp\E^+\n &= \beta_{11}/2\\
  \zeta^d - \zeta^+ &= -\gamma_2^+\beta_{10} -\gamma_1^+\beta_{13} \label{rhs_d}\\
  \n\cdot\vvs^d - \n\cdot\vvs^+ &= -(\cpII)^+\beta_{10} - (\cpI)^+\beta_{13} \label{rhs_e}\\
  \s\cdot\vvs^d - \s\cdot\vvs^+ &= -c_s^+\beta_{12} \label{rhs_f}\\
  \hatt\cdot\vvs^d - \hatt\cdot\vvs^+ &= -c_s^+\beta_{11}\\
  \n\cdot\vvf^d - \n\cdot\vvf^+ &= -(\gamma_2 \cpII)^+\beta_{10} - (\gamma_1 \cpI)^+\beta_{13} \label{rhs_g}\\
  \s\cdot\vvf^d - \s\cdot\vvf^+ &= (c_s \rhof/m)^+\beta_{12}\\
  \hatt\cdot\vvf^d - \hatt\cdot\vvf^+ &= (c_s \rhof/m)^+\beta_{11}.
\end{align}
Using the continuity condition (\ref{cont_cond_a}), (\ref{lhs_e}) and
(\ref{rhs_e}) we obtain

\begin{equation}
  \label{beta_sys1}
  (\cpI)^- \beta_1  + (\cpII)^- \beta_4 - (\cpII)^+\beta_{10}  - (\cpI)^+\beta_{13} = \n\cdot(\vvs^- - \vvs^+).
\end{equation}
Next from (\ref{cont_cond_c}), (\ref{lhs_g}) and (\ref{rhs_g}) we
obtain
\begin{equation}
  \label{beta_sys2}
  (\gamma_1 \cpI)^- \beta_1  + (\gamma_2 \cpII)^- \beta_4 - (\gamma_2 \cpII)^+\beta_{10}  - (\gamma_1 \cpI)^+\beta_{13} = \n\cdot(\vvf^- - \vvf^+).
\end{equation}

Using the continuity condition (\ref{cont_cond_d}) and the identity
(\ref{iden_1}) we obtain

\begin{equation}
  2\mufr^- \n\tp\E^c\n + \lambda^- (\n\tp\E^c\n + \s\tp\E^c\s  + \hatt\tp\E^c\hatt) -\alpha^- M^-\zeta^c
  = 2\mufr^+ \n\tp\E^d\n + \lambda^+ (\n\tp\E^d\n + \s\tp\E^d\s  + \hatt\tp\E^c\hatt) -\alpha^+ M^+\zeta^d
\end{equation}

We now substitute for $\E^c$ and $\E^d$ using (\ref{lhs_a}),
(\ref{lhs_b}), (\ref{lhs_d}), (\ref{rhs_a}), (\ref{rhs_b}) and
(\ref{rhs_d})

\begin{multline}
  \label{beta_sys3}
  (2\mufr^- + \lambda^- + \alpha^-M^-\gamma_1^-)\beta_1 
  + (2\mufr^- + \lambda^- + \alpha^-M^-\gamma_2^-)\beta_4
  + (2\mufr ^+ + \lambda^+ + \alpha^+M^+\gamma_2^+)\beta_{10} \\
  + (2\mufr^+ + \lambda^+ + \alpha^+M^+\gamma_1^+)\beta_{13} = \nt(\T^- - \T^+)\n.
\end{multline} 

Finally using the continuity condition (\ref{cont_cond_f}) and the
identity (\ref{iden_2}) we obtain
\begin{equation*}
  M^-\alpha^-(\n\tp\E^c\n + \s\tp\E^c\s) - M^-\zeta^c = M^+\alpha^+(\n\tp\E^d\n + \s\tp\E^d\s) - M^+\zeta^d.
\end{equation*}
Substituting again for $\E^c$ and $\E^d$ gives

\begin{equation}
  \label{beta_sys4}
  M^-(\alpha^- + \gamma_1^-)\beta_1 + M^-(\alpha^- +\gamma_2^-)\beta_4 + M^+(\alpha^+ +\gamma_2^+) \beta_{10} + M^+(\alpha^+ +\gamma_1^+)\beta_{13} = \nt(\Tf^- - \Tf^+)\n.
\end{equation}
There is no straightforward solution to the system
(\ref{beta_sys1})--(\ref{beta_sys4}). Inverting the coefficient matrix
$$
\begin{pmatrix}
  2\mufr^- + \lambda^- + \alpha^-M^-\gamma_1^- & 2\mufr^- + \lambda^- + \alpha^-M^-\gamma_2^- 
  &2\mufr^+ + \lambda^+ + \alpha^+M^+ \gamma_2^+ 
  &2\mufr^+ + \lambda^+ + \alpha^+M^+\gamma_1^+ \\
  M^-(\alpha^- +\gamma_1^-)  & M^-(\alpha^- +\gamma_2^-)& M^+(\alpha^+ +\gamma_2^+)& M^+(\alpha^+ +\gamma_1^+)\\
  (\cpI)^- & (\cpII)^- &-(\cpII)^+ &-(\cpI)^+\\
  (\gamma_1 \cpI)^- & (\gamma_2 \cpII)^- &-(\gamma_2 \cpII)^+ &-(\gamma_1 \cpI)^+
\end{pmatrix}
$$
we obtain the following expressions:
\begin{align}
  \beta_1 &= d_{11} \n\tp (\T^- - \T^+)\n + d_{12} \n\tp (\Tf^- - \Tf^+)\n 
  + d_{13}\n\cdot(\vvs^--\vvs^+)+d_{14}\n\cdot(\vvf^- - \vvf^+)\\
  \beta_4 &=d_{21} \n\tp (\T^- - \T^+)\n + d_{22} \n\tp (\Tf^- - \Tf^+)\n 
  + d_{23}\n\cdot(\vvs^--\vvs^+)+d_{24}\n\cdot(\vvf^- - \vvf^+) \\
  \beta_5 &=d_{31} \n\tp (\T^- - \T^+)\n + d_{32} \n\tp (\Tf^- - \Tf^+)\n 
  + d_{33}\n\cdot(\vvs^--\vvs^+)+d_{34}\n\cdot(\vvf^- - \vvf^+)\\
  \beta_8 &=d_{41} \n\tp (\T^- - \T^+)\n + d_{42} \n\tp (\Tf^- - \Tf^+)\n 
  + d_{43}\n\cdot(\vvs^--\vvs^+)+d_{44}\n\cdot(\vvf^- - \vvf^+)
\end{align}
Here the $d_{ij}$ are the entries of the inverse of the coefficient
matrix above.

Now we deal with the shear waves. Using the continuity condition
(\ref{cont_cond_e}) with the identity (\ref{iden_1b})
\begin{equation}
  2\mufr^-\s\tp \E^c\n = 2\mufr^+\s\tp \E^d\n.
\end{equation}
Substituting for $\E^c$ and $\E^d$ using (\ref{lhs_c}) and (\ref{rhs_c})
\begin{equation}
  (\mufr)^- \beta_2 + (\mufr)^+\beta_{12} = \s\tp(\T^- - \T^+)\n.
\end{equation}
Finally using (\ref{cont_cond_b}), (\ref{lhs_f}) and (\ref{rhs_f}) gives
\begin{equation}
  (c_s)^-\beta_2 - (c_s)^+ \beta_{12} = \s\cdot(\vvs^- -\vvs^+).
\end{equation}
Therefore,
\begin{align}
  \beta_2 &= \frac{(c_s)^+ \s\tp(\T^- - \T^+)\n + \mufr^+ \s\cdot(\vvs^- -\vvs^+)}{(c_s)^+(\mufr)^- + (c_s)^-(\mufr)^+} \\
  \beta_{12} &= \frac{(c_s)^- \s\tp(\T^- - \T^+)\n - \mufr^- \s\cdot(\vvs^- -\vvs^+)}{(c_s)^+(\mufr)^- + (c_s)^-(\mufr)^+}.
\end{align}
In a similar manner using the continuity relationships (\ref{cont_cond_e}) we obtain

\begin{align}
  \beta_3 &= \frac{(c_s)^+ \hatt\tp(\T^- - \T^+)\n + \mufr^+ \hatt\cdot(\vvs^- -\vvs^+)}{(c_s)^+(\mufr)^- + (c_s)^-(\mufr)^+} \\
  \beta_{11} &= \frac{(c_s)^- \hatt\tp(\T^- - \T^+)\n - \mufr^- \hatt\cdot(\vvs^- -\vvs^+)}{(c_s)^+(\mufr)^- + (c_s)^-(\mufr)^+}.
\end{align}

\subsection{Upwind numerical flux}
\label{sec:flux_poro}

We define an upwind numerical flux $(\Pi\q)^*$ along $\n$ by

\begin{equation}
  (\Pi\q)^* = \Pi^-\q^- +Q^{-}(\beta_1(\cpI)^-\rr_1^- + \beta_2(c_s)^-\rr_2^- + \beta_3(c_s)^-\rr_3^- 
  +\beta_4(\cpII)^-\rr_4^-).
\end{equation}
We now compute the $\beta_i \rr_i$ terms.  First, noting that $\rr_i = L \rr_i^\prime$, a simple computation gives

$$
\rr_1^- =
\begin{pmatrix}
  \overline{\n\otimes\n} \\
  -\gamma_1^-\\ 
  (\cpI)^-\n\\
  \gamma_1^- (\cpI)^-\n
\end{pmatrix},
\quad 
\rr_2^- =
\begin{pmatrix}
  \overline{\n\otimes\s} \\
  0\\ 
  (c_s)^-\s\\
  -\frac{(c_s)^- \rhof^-}{m^-}\s
\end{pmatrix},
\quad %
\rr_3^- =
\begin{pmatrix}
  \overline{\n\otimes\hatt} \\
  0\\ 
  (c_s)^-\hatt\\
  -\frac{(c_s)^- \rhof^-}{m^-}\hatt
\end{pmatrix},
\quad
\rr_4^- =
\begin{pmatrix}
  \overline{\n\otimes\n} \\
  -\gamma_2^-\\ 
  (\cpII)^-\n\\
  \gamma_2^- (\cpII)^-\n
\end{pmatrix},
$$
where $\overline{\n\otimes\n} = (n_1^2, n_2^2, n_3^2, n_1 n_2, n_2
n_3, n_1
n_3)\tp$ is a flattened representation of the tensor
$\n\otimes\n$, etc.

In what follows, we make
multiple use of the vector/tensor identities

\begin{align}
(\s\cdot\mathbf{a})\s + (\hatt\cdot\mathbf{a})\hatt  &= -\n\times(\n\times\mathbf{a})\\
(\s\cdot\mathbf{a})\sym(\s\otimes\n) + (\hatt\cdot\mathbf{a})\sym(\hatt\otimes\n)  &= -\sym(\n\otimes(\n\times(\n\times\mathbf{a})))
\end{align}

We define
\begin{align*}
  \sbb{\T} &=  \T^- \n^- + \T^+\n^+\\
  \sbb{\Tf} &=  \Tf^- \n^- + \Tf^+\n^+\\
  \sbb{\vv} &= \n^{-}{}\tp \vv^- + \n^{+}{}\tp \vv^+\\
  \ssb{\vv} &= \vv^- - \vv^+
\end{align*}
For the fast P-wave term we have
\begin{equation}
  \beta_1 (\cpI)^- \rr_1^- =
  (\cpI)^-(d_{11} \nt\sbb{\T} + d_{12} \nt\sbb{\Tf} + d_{13}\sbb{\vvs} + d_{14}\sbb{\vvf}) \times\\
\begin{pmatrix}
  \overline{\n\otimes\n} \\
  -\gamma_1^-\\ 
  (\cpI)^-\n\\
  \gamma_1^- (\cpI)^-\n
\end{pmatrix},
\end{equation}
For the S-wave term we have
\begin{multline}
  \beta_2 c_s^- \rr_2^- +  \beta_3 c_s^- \rr_3^- =
     \frac{-(c_s)^-(c_s)^+}{ (c_s)^+(\mufr)^- + (c_s)^-(\mufr)^+ }
\begin{pmatrix}
\overline{\sym(\n\otimes(\n\times(\n\times\sbb{\T})))}\\
0\\
(c_s)^- \n\times(\n\times\sbb{\T})\\
\frac{-(c_s)^- \rho_f^-}{m^-} \n\times(\n\times\sbb{\T})
\end{pmatrix}
\\ %
- \frac{(c_s)^-(\mufr)^+}{ (c_s)^+(\mufr)^- + (c_s)^-(\mufr)^+ }
\begin{pmatrix}
\overline{\sym(\n\otimes(\n\times(\n\times\sbb{\vvs})))}\\
0\\
(c_s)^- \n\times(\n\times\ssb{\vvs})\\
\frac{-(c_s)^- \rho_f^-}{m^-} \n\times(\n\times\ssb{\vvs})
\end{pmatrix}
\end{multline}

Finally for the slow P-wave we have
\begin{equation}
  \beta_3 (\cpII)^- \rr_1^- =
  (\cpII)^-(d_{21} \nt\sbb{\T} + d_{22} \nt\sbb{\Tf} + d_{23}\sbb{\vvs} + d_{24}\sbb{\vvf}) \times\\
\begin{pmatrix}
  \overline{\n\otimes\n} \\
  -\gamma_2^-\\ 
  (\cpII)^-\n\\
  \gamma_2^- (\cpII)^-\n
\end{pmatrix} 
\end{equation}

\section{Consideration of poro-viscoelasticity}
\label{sec:dissip}

\subsection{Introduction}

The low-frequency regime is straightforward and follows Biot's 1956
paper \cite{biot56a}.  Using the conventions of equations
(\ref{eq:biot2a}) and (\ref{eq:biot2b}), the low-frequency dissipative
regime is modelled by the term
$\displaystyle{\frac{\eta}{k}\pd{\w}{t}}$.  For the hyperbolic system
(\ref{hyp_sys_mat}) we simply add the source term
(\ref{low_freq_source}).  We note that in certain physical situations
(when the permeability of the solid matrix is very small and the
frequency content of the propagating wave very low) the second P-wave
can be essentially static and highly diffusive (so has a
characteristic timescale much smaller than the time step of the
non-dissipative hyperbolic system), rendering the system stiff and
requiring extremely small time steps in an explicit scheme to capture
the dissipative effects.  This is considered by Carcione and
Quiroga-Goode in \cite{carcione95} who used an operator splitting
approach to avoid this issue and treated the viscous dissipation term
analytically.  In a more recent paper Lemoine et al.  \cite{lemoine13}
work in a finite volume setting and again implement an operator
splitting on the dissipative part, while an IMEX scheme is implemented
in \cite{dwle17}. Here we consider both operator-splitting and IMEX
techniques; see Section~\ref{sec:numer-exper} below.

\subsection{High-frequency case}
\label{high_freq}

In the high-frequency case the term
$\displaystyle{\frac{\eta}{k}\pd{\w}{t}}$ in
equation~(\ref{eq:biot2b}) is replaced by a convolution
$\displaystyle{b*\pdd{\w}{t}}$ where $b(t) = \frac{\eta}{k} \Psi(t)
H(t)$, $\Psi(t)$ is a relaxation function of the form

\begin{equation}
  \Psi(t) = 1 + \sum_{l=1}^L \left(\frac{\tau_\epsilon^l}{\tau_\sigma^l} - 1\right) e^{-t/\tau_\sigma^l}
\end{equation}
with relaxation times $\tau_\epsilon$ and $\tau_\sigma$, and $H(t)$ is
a Heaviside function.  Thus the relaxation mechanism corresponds to a
generalised Zener model; see \cite{carcione15}.  In practice it is
common to deal with a single Zener model, which is the case we deal
with here.  We have
\begin{align}
  b*\pd{\vvf}{t}  &= \frac{\eta}{k} \int_{-\infty}^t \Psi (t-\tau) \pd{\vvf}{\tau} d\tau\\
  &= \frac{\eta}{k} \int_{-\infty}^t  \pd{\vvf}{\tau} d\tau + \frac{\eta}{k} \sum_{l=1}^L    
  \left(\frac{\tau_\epsilon^l}{\tau_\sigma^l} - 1\right) 
  \int_{-\infty}^t e^{-(t-\tau)/\tau_\sigma^l}  \pd{\vvf}{\tau} d\tau\\
  &= \frac{\eta}{k} \vvf + \frac{\eta}{k} \sum_{l=1}^L    
  \left(\frac{\tau_\epsilon^l}{\tau_\sigma^l} - 1\right) 
  \int_{-\infty}^t e^{-(t-\tau)/\tau_\sigma^l}  \pd{\vvf}{\tau} d\tau
\end{align}
Introducing memory variables
\begin{equation}
  \ee^l = \left(\frac{\tau_\epsilon^l}{\tau_\sigma^l} - 1\right) 
  \int_{-\infty}^t e^{-(t-\tau)/\tau_\sigma^l}  \pd{\vvf}{\tau} d\tau
\end{equation}
we obtain $3L$ additional differential equations:
\begin{align}
  \pd{\ee^l}{t} &= \left(\frac{\tau_\epsilon^l}{\tau_\sigma^l} - 1\right)  \pd{\vvf}{t}  - \frac{\ee^l}{\tau_\sigma^l}
  \label{mem_eq}
\end{align}
and
\begin{equation}
  b*\pd{\vvf}{t} = \frac{\eta}{k}\vvf + \frac{\eta}{k}\sum_{l=1}^L \ee^l.
\end{equation}
It is customary to express the relaxation times in terms of a
quality factor $Q_0$ and a reference frequency $f_0$ as
\begin{align}
  \tau_\epsilon &= (\sqrt{Q_0^2 +1} +1)/(2\pi f_0 Q_0) \\
  \tau_\sigma   &= (\sqrt{Q_0^2 +1} -1)/(2\pi f_0 Q_0).
\end{align}
For $L=1$ the variable $\q$ defined in (\ref{hypsys_cf}) must now be
augmented with three additional variables $e^1_x, e^1_y, e^1_z$:
\begin{equation}
  \label{hypsys_cf_highfreq}
  \q = (\epsilon_{11}, \epsilon_{22},\epsilon_{33},\epsilon_{12},\epsilon_{23}, \epsilon_{13},\zeta, \us, \vs, \ws, \uf, \vf, \wf, e^1_x, e^1_y, e^1_z)\tp
\end{equation}
and the various coefficient matrices inflated in an obvious manner.

As noted in \cite{dwle17} implementation of the high-frequency case
needs to be carried out with some care.  Solving the sixteen-variable
system as an inflated hyperbolic system results in a memory variable
that converges to zero very quickly.  An accurate scheme is obtained
by treating the memory equations (\ref{mem_eq}) as an uncoupled system
of ordinary differential equations and evaluating
$\displaystyle{\pd{\vvf}{t}}$ from its gradient and flux terms.

\section{Elastic/poroelastic coupling}\label{sec:elast-coupl}

In many applications to geophysics, one is interested in coupling
elastic and poroelastic wave propagation; see \cite{lahivaara14,
  lahivaara15, lahivaara18}.  In this section we outline the DG discretisation for
three-dimensional elastic waves for an isotropic medium, again for a
velocity/strain formulation.  This results for the elastic case were
given in \cite{wilcox10}, and are simply summarised below for
convenience and consistency with the conventions of this paper. We
then derive numerical fluxes for the interface between elastic and
poroelastic elements.

Expressed as a second-order system the elastic wave equation takes the form

\begin{equation}
  \rhoe\pdd{\uue}{t} = \nabla\cdot \boldS
\end{equation}
  where $\rhoe$ is density and $\boldS$ is a stress tensor.  In the isotropic case we
  consider here $\boldS$ may be written in the usual form

\begin{equation}
  \boldS = 2\mue {\bf E} + \lambdae \trace({\bf E}){\bf I} 
\end{equation}
where ${\bf E}$ is the solid strain tensor and $\mue$ and $\lambdae$
are Lam{\'e} coefficients. Expressed as a first-order hyperbolic
system with variable
\begin{equation}
  \label{hypsys_ew}
  \q_e = (\epsilon_{11}, \epsilon_{22}, \epsilon_{33}, \epsilon_{12}, \epsilon_{23}, \epsilon_{13}, u_{\rm e},
  v_{\rm e}, w_{\rm e})\tp
\end{equation}
where $\vve=(u_{\rm e},v_{\rm e}, w_{\rm e})$ are the $x$, $y$ and $z$ components of
the velocity $\displaystyle{\pd{\uue}{t}}$ gives
\begin{equation}
  Q_e \pd{\q}{t} + \nabla\cdot\mathcal{F}_e 
  = Q_e \pd{\q}{t} + \pd{(A_e^i \q)}{x_i} =0 %
\end{equation}
where
$$
\mathcal{F}_e = [F_1, F_2, F_3] = [A_e^1 \q, A_e^2 \q, A_e^3 \q],
$$
and

\begin{equation}
  Q_e =
\blockmatrix{I}{0}{0}{Q_{e,2}}
\end{equation}
(here $I$ is the $6\times 6$ identity matrix) and

\begin{equation}
  Q_{e,2} =
  \begin{pmatrix}
    \rhoe &0 &0\\
    0     &\rhoe &0\\
    0     &0 &\rhoe
  \end{pmatrix}.
\end{equation}
As in equation \eqref{eq:jac_mat} the off-diagonal blocks of the Jacobian matrices $A_e^i$, $i=1,2,3$
are given in Table~\ref{tab:Jacobian-block-elastic}.
\begin{table}
\caption{The off-diagonal blocks of the Jacobian matrices $A_e^i$.}
\label{tab:Jacobian-block-elastic}
\begin{align*}
  A_{e,1}^1 &= - \begin{pmatrix}
    2\mue + \lambdae & \lambdae & \lambdae  & 0     &0 &0    \\
                   0 &        0 &        0  & 2\mue &0 &0    \\
                   0 &        0 &        0  & 0     &0 &2\mue    
                 \end{pmatrix},
        & A_{e,2}^1 = - \begin{pmatrix}
    1 &0 &0\\
    0 &0 &0\\
    0 &0 &0\\
    0 &1/2 &0\\
    0 &0   &0\\
    0 &0 &1/2\\ 
  \end{pmatrix}\\
   A_{e,1}^2 &= - \begin{pmatrix}
    0 & 0 & 0 & 2\mue & 0 & 0\\
    \lambdae & 2\mue + \lambdae &\lambdae & 0 & 0 & 0\\
    0 & 0 & 0 & 0 & 2\mue & 0\\
  \end{pmatrix}
      & A_{e,2}^2 = - \begin{pmatrix}
    0   & 0 & 0\\
    0   & 1 & 0\\
    0   & 0 & 0\\
    1/2 & 0 & 0\\
    0   & 0 & 1/2\\
    0   & 0 & 0\\
  \end{pmatrix}\\
  A_{e,1}^3 &= - \begin{pmatrix}
    0 & 0 & 0 & 0 & 0     & 2\mue\\
    0 & 0 & 0 & 0 & 2\mue & 0 \\
    \lambdae & \lambdae & 2\mue + \lambdae  & 0 & 0 & 0\\
  \end{pmatrix}
      & A_{e,2}^2 = - 
  \begin{pmatrix}
    0   & 0 & 0\\
    0   & 0 & 0\\
    0   & 0 & 1\\
    0   & 0 & 0\\
    0   & 1/2 & 0\\
    1/2 & 0 & 0\\
  \end{pmatrix}
\end{align*}
\end{table}

We have the well-known expressions for elastic wave speeds
\begin{equation}
  c_p = \pm\sqrt{\frac{\lambdae+2\mue}{\rhoe}} \quad\text{and}\quad c_s = \pm\sqrt{\frac{\mue}{\rhoe}}.
\end{equation}
Solving the Riemann problem as before we obtain the following coefficients corresponding to the non-zero wave speeds
\begin{align}
  \beta_1 &= \frac{(c_p)^+ \n\tp(\boldS^- - \boldS^+)\n + (\lambdae^+ + 2\mue^+) \n\cdot(\vve^- -\vve^+)}{(c_p)^+(\lambdae^- +2\mue^-) + (c_p)^-(\lambdae^+ +2\mue^+)} \\
  \beta_2 &= \frac{(c_s)^+ \s\tp(\boldS^- - \boldS^+)\n + \mue^+ \s\cdot(\vve^- -\vve^+)}{(c_s)^+(\mue)^- + (c_s)^-(\mue)^+} \\
  \beta_3 &= \frac{(c_s)^- \hatt\tp(\boldS^- - \boldS^+)\n + \mue^- \hatt\cdot(\vve^- -\vve^+)}{(c_s)^+(\mue)^- + (c_s)^-(\mue)^+}\\
\end{align}

Defining an upwind numerical flux $(\Pi\q)^*$ along $\n$ by
\begin{equation}
  (\Pi\q)^* = \Pi^-\q^- +Q^{-}(\beta_1(c_p)^-\rr_1^- + \beta_2(c_s)^-\rr_2^- + \beta_3(c_s)^-\rr_3^-)
\end{equation}
where

$$
\rr_1^- = 
\begin{pmatrix}
\overline{\n\otimes\n} \\
(c_p)^- \n
\end{pmatrix},
\quad 
\rr_2^- =
\begin{pmatrix}
\overline{\n\otimes\s} \\
(c_s)^- \s
\end{pmatrix},
\quad %
\rr_3^- =
\begin{pmatrix}
\overline{\n\otimes\hatt} \\
(c_s)^- \hatt
\end{pmatrix},
$$
where $\n = (\hn_1, \hn_2, \hn_3)\tp$, $\s = (\hs_1, \hs_2, \hs_3)\tp$ and $\ttt = (\hht_1, \hht_2, \hht_3)\tp$.

We define
\begin{align*}
  \sbb{\boldS} &=  \boldS^- \n^- + \boldS^+\n^+\\
  \sbb{\vve} &= \n^{-}{}\tp \vve^- + \n^{+}{}\tp \vve^+\\
  \ssb{\vve} &= \vve^- - \vve^+,
\end{align*}
and obtain an upwind flux

\begin{multline}
  \beta_1 (c_p)^- \rr_1^- + \beta_2 c_s^- \rr_2^- +  \beta_3 c_s^- \rr_3^-  = \\
      \frac{(c_p)^- c_p^+ \nt\sbb{\boldS} + (c_p)^- (\lambdae^+ + 2\mue^+)\sbb{\vve}}{c_p^+(\lambdae^-+2\mue^-) + c_p^-(\lambdae^++2\mue^+)}
       \begin{pmatrix}
         \overline{\n\otimes\n} \\
         (c_p)^- \n
       \end{pmatrix}\\
      -\frac{(c_s)^-(c_s)^+}{ (c_s)^+(\mue)^- + (c_s)^-(\mue)^+ }
       \begin{pmatrix}
         \overline{\sym(\n\otimes(\n\times(\n\times\sbb{\T})))}\\
         (c_s)^- \n\times(\n\times\sbb{\T})
       \end{pmatrix}\\
      - \frac{(c_s)^-(\mue)^+}{ (c_s)^+(\mue)^- + (c_s)^-(\mue)^+ }
       \begin{pmatrix}
         \overline{\sym(\n\otimes(\n\times(\n\times\sbb{\vve})))}\\
         (c_s)^- \n\times(\n\times\ssb{\vvs})
       \end{pmatrix}
\end{multline}

\subsection{Elastic/poroelastic interface}

\begin{figure}[!h]
  \centering
  \includegraphics[width=0.49\textwidth]{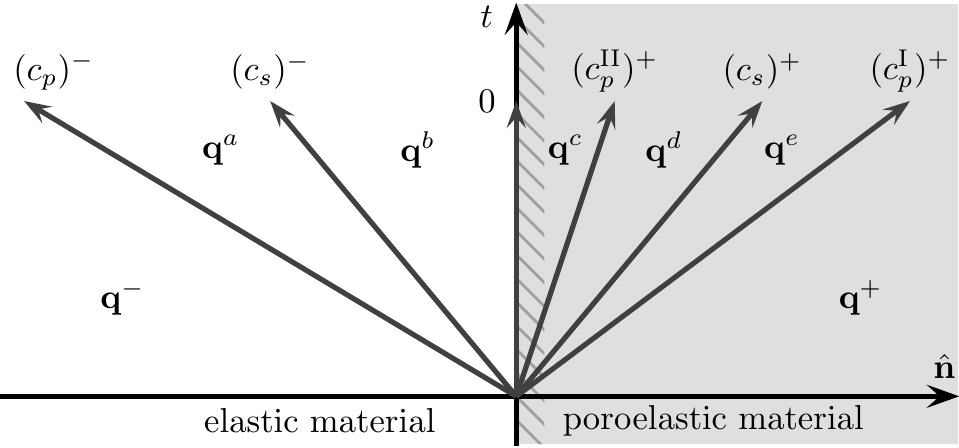}

  \caption{Schematic showing characteristic wave speeds at an
    elastic/poroelastic interface between two states $\q^-$ (elastic)
    and $\q^+$ (poroelastic). $\q^a$-- $\q^e$ denote the intermediate
    states.}
  \label{fig:riemann_prob_el_poro_int}
\end{figure}

As in \cite{dwle17} we solve a Riemann problem at the interface
subject to the following flux continuity conditions at the interface:

\begin{align}
  \n\cdot\vve^b &= \n\cdot\vvs^c\label{cont_cond_aa}\\
  \s\cdot\vve^b &= \s\cdot\vvs^c\label{cont_cond_bb}\\
  \hatt\cdot\vve^b &= \hatt\cdot\vvs^c\\
  0  &= \n\cdot\vvf^c\label{cont_cond_cc}\\
  \n\tp \boldS^b \n &= \n\tp \T^c \n \label{cont_cond_dd}\\
  \s\tp\boldS^b\n &= \s\tp\T^c\n\label{cont_cond_ee}\\
  \hatt\tp\T^c\n &= \hatt\tp\T^d\n%
\end{align}
where we now have 7 unknown states shown in Figure~\ref{fig:riemann_prob_el_poro_int}.

Note that the normal fluid and solid velocities in the poroelastic
medium are assumed to be the same as the solid velocity in the elastic
medium at the interface.  From the Rankine-Hugoniot conditions we
obtain

\begin{align}
  \q^- - \q^a &= \beta_1^e \rr_1^e \\
  \q^a - \q^b &= \beta_2^e \rr_2^e + \beta_3^e \rr_3^e \\
  \q^c - \q^d &= \beta_{10}^p \rr_{10}^p \\
  \q^d - \q^e &= \beta_{11}^p \rr_{11}^p + \beta_{12}^p \rr_{12}^p \\
  \q^e - \q^-&= \beta_{13}^p \rr_{13}^p 
\end{align}
where $\rr_j^e$ is an eigenvector for the elastic domain and $\rr_j^p$
is an eigenvector for the poroelastic domain corresponding to wave
speeds $c_j^\pm$ and hence

\begin{align}
  \q^- - \q^b &= \beta_1^e \rr_1^e + \beta_2^e \rr_2^e \label{link_1_ep}\\
  \q^c - \q^+ &= \beta_{10}^p \rr_{10}^p + \beta_{11}^p \rr_{11}^p +\beta_{12}^p \rr_{12}^p +\beta_{13}^p \rr_{13}^p. \label{link_2_ep}
\end{align}
Using (\ref{cont_cond_aa}), (\ref{cont_cond_cc}) and (\ref{cont_cond_dd}) we obtain

\begin{align}
  (c_p^e)^- \beta_1^e  - (\cpII)^+\beta_{10}^p  - (\cpI)^+\beta_{13}^p &= \n\cdot(\vve^- - \vvs^+)\\
  (\gamma_2 \cpII)^+\beta_{10}^p  + (\gamma_1  \cpI)^+\beta_{13}^p &= \n\cdot\vvf^+\\
  (2\mue^- + \lambdae^- )\beta_1^e 
  + (2\mufr^+ + \lambda^+ + \alpha^+M^+\gamma_2^+)\beta_{10} 
  + (2\mufr^+ + \lambda^+ + \alpha^+M^+\gamma_1^+)\beta_{13} &= \nt(\boldS^- - \T^+)\n.
\end{align}
As in the poroelastic case, we invert the coefficient matrix
$$
\begin{pmatrix}
  2\mue^- + \lambdae^-  &2\mufr^+ + \lambda^+ + \alpha^+M^+\gamma_2^+ &2\mufr^+ + \lambda^+ + \alpha^+M^+\gamma_1^+\\
  (c_p^e)^- &- (\cpII)^+  &-(\cpI)^+\\
  0        & (\gamma_2 \cpII)^+  &(\gamma_1  \cpI)^+
\end{pmatrix}
$$
to solve for $\beta_1^e$, $\beta_{10}^p$ and $\beta_{13}^p$ and obtain coefficients $\tilde{d}_{ij}$ such that

\begin{align}
  \beta_1^e &= \tilde{d}_{11} \n\tp (\boldS^- - \T^+)\n + \tilde{d}_{12}\n\cdot(\vve^--\vvs^+)\ + \tilde{d}_{13}\n\cdot\vvf^+\\
  \beta_{10}^p &= \tilde{d}_{21} \n\tp (\boldS^- - \T^+)\n + \tilde{d}_{22}\n\cdot(\vve^--\vvs^+)\ + \tilde{d}_{23}\n\cdot\vvf^+\\
  \beta_{13}^p &= \tilde{d}_{31} \n\tp (\boldS^- - \T^+)\n + \tilde{d}_{32}\n\cdot(\vve^--\vvs^+)\ + \tilde{d}_{33}\n\cdot\vvf^+.
\end{align}

Finally, we deal with the shear waves. Using (\ref{cont_cond_bb}) and
(\ref{cont_cond_ee}) we obtain
\begin{align}
  \mue \beta_2^e + \mufr^p\beta_{12}^p &= \s\tp(\boldS^- - \T^+)\n\\
  (c_s^e)^-\beta_2^e - (c_s^p)^+ \beta_{12}^p &= \s\cdot(\vve^- -\vvs^+).
\end{align}

Therefore,
\begin{align}
  \beta_2^e &= \frac{(c_s^p)^+ \s\tp(\boldS^- - \T^+)\n + \mufr^+ \s\cdot(\vve^- -\vvf^+)}{(c_s^p)^+(\mue)^- + (c_s^e)^-(\mufr)^+} \\
  \beta_{12}^p &= \frac{(c_s^e)^- \s\tp(\boldS^- - \T^+)\n - \mue^- \s\cdot(\vve^- -\vvs^+)}{(c_s^p)^+(\mue)^- + (c_s^e)^-(\mufr)^+}.
\end{align}

Similarly we obtain

\begin{align}
  \beta_3^e &= \frac{(c_s^p)^+ \hatt\tp(\boldS^- - \T^+)\n + \mufr^+ \hatt\cdot(\vve^- -\vvf^+)}{(c_s^p)^+(\mue)^- + (c_s^e)^-(\mufr)^+} \\
  \beta_{11}^p &= \frac{(c_s^e)^- \hatt\tp(\boldS^- - \T^+)\n - \mue^- \hatt\cdot(\vve^- -\vvs^+)}{(c_s^p)^+(\mue)^- + (c_s^e)^-(\mufr)^+}.
\end{align}

\subsection{Upwind numerical flux}

For the interface element on the elastic domain, we define an upwind
numerical flux $(\Pi\q)^*$ along $\n$ by

\begin{equation}
  (\Pi\q^e)^* = \Pi^-\q^- +Q^{-}(\beta_1^e(c_p^e)^-\rr_1^- + \beta_2^e(c_s^e)^-\rr_2^- + \beta_3^e(c_s^e)^-\rr_3^- )
\end{equation}
while, for the poroelastic domain, we define an upwind numerical flux
$(\Pi\q)^*$ along $\n$ by

\begin{equation}
  (\Pi\q^p)^* = \Pi^+\q^+ - Q^+(\beta_{10}^p(\cpII)^+\rr_{10}^+ + \beta_{11}^p(c_s)^+\rr_{11}^+ + \beta_{12}^p(c_s)^+\rr_{12}^+ + \beta_{13}^p(\cpI)^+\rr_{13}^+ ).
\end{equation}
We define
\begin{align*}
  \sbb{\boldS,\T} &=  \boldS^- \n^- + \T^+\n^+\\
  \sbb{\vve, \vvs} &= \n^{-}{}\tp \vve^- + \n^{+}{}\tp \vvs^+\\
  \sbb{\vvf} &= \n^{+}{}\tp \vvf^+\\
  \ssb{\vve,\vvs} &= \vve^- - \vvs^+
\end{align*}
We now assemble the flux terms for the elastic element:

\begin{multline}
  \beta_1^e (c_p^e)^- \rr_1^{-,e}  +\beta_2 c_s^- \rr_2^- +  \beta_3 c_s^- \rr_3^-  = \\
              (c_p^e)^-(d_{11} \nt\sbb{\boldS,\T} + d_{12} \sbb{\vve,\vvs} + d_{13}\sbb{\vvf} ) \times
                \begin{pmatrix}
                  \overline{\n\otimes\n} \\
                  (c_p)^- \n
                \end{pmatrix}\\
               -\frac{(c_s^e)^-(c_s)^+}{ (c_s)^+(\mue)^- + (c_s^e)^-(\mufr)^+ }
                 \begin{pmatrix}
                   \overline{\sym(\n\otimes(\n\times(\n\times\sbb{\boldS, \T})))}\\
                   (c_s)^- \n\times(\n\times\sbb{\boldS,\T})
                 \end{pmatrix}
                 \\ %
                 - \frac{(c_s^e)^-(\mufr)^+}{ (c_s)^+(\mue)^- + (c_s)^-(\mufr)^+ }
                   \begin{pmatrix}
                     \overline{\sym(\n\otimes(\n\times(\n\times\ssb{\vve, \vvs})))}\\
                     (c_s)^- \n\times(\n\times\ssb{\vve,\vvs})
                   \end{pmatrix}
\end{multline}

Finally, we assemble the flux terms for the poroelastic element:

\begin{multline}
  \beta_{10}^p (\cpII)^+ \rr_{10}^+ + \beta_{11}^p (c_s^p)^+ \rr_{11}^+ + \beta_{12}^p (c_s^p)^+ \rr_{12}^+ +  \beta_{13}^p (\cpI)^+ \rr_{13}^+ =\\
     (\cpII)^+(\tilde{d}_{21} \nt\sbb{\boldS,\T} + \tilde{d}_{22} \sbb{\vve,\vvs} + \tilde{d}_{23}\sbb{\vvf} )
       \begin{pmatrix}
         \overline{\n\otimes\n} \\
         -\gamma_2^+\\ 
         -(\cpII)^+\n\\
         -\gamma_2^+ (\cpII)^+\n
       \end{pmatrix}\\
       \frac{-1}{(c_s^p)^+(\mue)^- + (c_s^e)^-(\mufr)^+} \times
        \left\{
          (c_s^e)^-(c_s^p)^+
          \begin{pmatrix}
            \overline{\sym(\n\otimes(\n\times(\n\times\sbb{\boldS,\T})))}\\
            0\\
            -(c_s)^- \n\times(\n\times\sbb{\boldS,\T})\\
            \frac{(c_s)^- \rho_f^-}{m^-} \n\times(\n\times\sbb{\boldS,\T})
          \end{pmatrix}
        \right .
        \\ %
        \left .
          - \mue^- (c_s^p)^+ 
          \begin{pmatrix}
            \overline{\sym(\n\otimes(\n\times(\n\times\ssb{\vve, \vvs})))}\\
            0\\
            -(c_s)^- \n\times(\n\times\ssb{\vve,\vvs})\\
            \frac{(c_s)^- \rho_f^-}{m^-} \n\times(\n\times\ssb{\vve,\vvs})
          \end{pmatrix}
        \right\}.\\
        (\cpI)^+(\tilde{d}_{31}\nt\sbb{\boldS,\T} + \tilde{d}_{32} \sbb{\vve,\vvs} + \tilde{d}_{33}\sbb{\vvf} ) \times%
        \begin{pmatrix}
   \overline{\n\otimes\n} \\
   -\gamma_1^+\\ 
   -(\cpI)^+\n\\
   -\gamma_1^+ (\cpI)^+\n
 \end{pmatrix}
\end{multline}

\section{Adjoint method}
\label{sec:adjoint}

In applications to inverse problems, we wish to quantify a model's fit
to observed data. In seismic problems data normally consists of ground
motion measurements following a seismic event due to a passive or
active source. Here we are interested in fitting full waveform ground
acceleration or velocity data, which requires simulating a forward
model many times. Poroelastic wave inverse problems are particularly
challenging since most nontrivial problems require multiparameter
estimation and the computational cost of the forward problem is
expensive and often prohibitive \cite{lahivaara14, lahivaara15}.  In
both frequentist and Bayesian approaches to inverse problems, a least
squares estimate is a good starting point to solving an inverse
problem. This requires the solution of a PDE-constrained optimisation
problem.

We introduce the following notation

\begin{equation}
  \label{forward}
  L(\q) = Q\pd{\q}{t} +\pd{}{x_i}(A^i \q)
\end{equation}
where
$\q = (\epsilon_{11}, \epsilon_{22},\epsilon_{33}, \epsilon_{12},
\epsilon_{23}, \epsilon_{13},\zeta, \us, \vs, \ws, \uf, \vf,\wf)\tp$
and we assume the Einstein summation convention over repeated
indices. We consider the hyperbolic system $L(\q)=\g+\g_V $.

Given time-varying data $\boldd(x_r,t)$ define the misfit functional

\begin{equation}
  \label{missfit}
  \chi(\boldtheta) = \frac{1}{2} \sum_{i\in\mathcal{I}, r\in\mathcal{R}}\int_0^T\int_\Omega
            [q_i(\boldtheta,x,t) - d_i(x_r,t)]^2 \delta(x-x_r) dxdt
\end{equation}
where $\q(\boldtheta, x,t)$ is the forward map evaluated on the
parameter set $\boldtheta$, $\mathcal{I}$ is an index set over the
observed measurements (i.e. which components of $\q$ are measured,
usually velocities), and $\mathcal{R}$ is an index set over the receiver
locations $x_r$. Gradient-based approaches to minimising
(\ref{missfit}) require estimation of the Jacobian of (\ref{missfit})
with respect to the parameter space $\boldtheta$ which usually
requires many evaluations of the forward map; this is an expensive
calculation as noted above. The adjoint method is a standard approach
for computing derivatives of a misfit functional in computational
seismology which reduces the number of evaluations of the forward map
to one together with one evaluation of a dual, or adjoint,
map. Fichtner gives an interesting history of the adjoint method in
seismology \cite{fichtner11}. When the elastic or poroelastic wave
equation is written as a second order system in time, the adjoint map
is self-adjoint, although time-reversed, which means the forward
solver can be used to solve the adjoint problem, and hence estimate
the Jacobian of the least squares misfit functional, \cite{tromp08},
\cite{fichtner11}.  With a first order system this is no longer the
case, and more care must be taken to both derive and solve an adjoint
equation. Since adjoints are not unique being specified relative to an
inner product, the actual choice of the inner product turns out to be
crucial to obtain an adjoint equation that is physically
meaningful. This was considered in \cite{wilcox15} for the elastic
wave equation.

We therefore replace (\ref{missfit}) by

\begin{equation}
  \label{missfit2}
  \chi(\boldtheta) = \frac{1}{2} \sum_{i\in\mathcal{I}, r\in\mathcal{R}}\int_0^T\int_\Omega
            [q_i(\boldtheta,x,t) - d_i(x,t)]^2 w_i \delta(x-x_r) dxdt
\end{equation}
where $w_i$ are positive weights.  We can then define an inner product $\langle \cdot, \cdot \rangle_W$ and write (\ref{missfit2}) as

\begin{equation}
  \label{eq:in_prod}
  \chi(\boldtheta) = \frac{1}{2} \langle \q - \boldd, (\q - \boldd)\delta(x-x_r) \chi_\mathcal{I}\rangle_W
\end{equation}
where $\chi_{\mathcal{I}}$ is an indicator function on the measurement
set (=1 if $q_i$ is measured, otherwise 0) and for simplicity we have
assumed just one receiver location.  In the following we take $w_i=1$
except for $i=4,5,6$ where we set $w_i=2$. The reason for this is that
we may write $\q$ in block form

\begin{equation}
  \label{eq:decomp}
  \q =
  \blockvector{\q_1}{\q_2}
\end{equation}
where $\q_1$ contains the 7 strain components
$\q_1 = (\epsilon_{11}, \epsilon_{22},\epsilon_{33}, \epsilon_{12},
\epsilon_{23}, \epsilon_{13},\zeta)\tp$, while $\q_2$ contains the 6
velocity components $\q_2 =( \us, \vs, \ws, \uf, \vf,\wf)\tp$. Note
that the first six entries of $\q_1$ is a flattened representation of
the strain tensor \eqref{eq:strain_tensor}, and the natural inner
product is given by the double dot product $\colon$, and the
off-diagonal (shear) terms are counted twice. Hence the specification
of weights above. In practice the inner product
$\langle\cdot, \cdot\rangle_W$ defined above makes no difference to
the estimation problem since ground motion data is measured and not
strain data.

In the following derivations we assume, for simplicity, that the
source parameters are known.  We define the directional derivative

\begin{equation}
  D_{\delta\boldtheta}\chi(\boldtheta) = \lim_{h\rightarrow 0}
       \frac{1}{h}[\chi(\boldtheta+h\delta\boldtheta) - \chi(\boldtheta)]
     \end{equation}
Then

\begin{equation}
\label{eq:dd_chi}
       D_{\delta\boldtheta}\chi = \left\langle (\q - \boldd)\delta(x-x_r)\chi_{\mathcal{I}}, D_{\delta\boldtheta}\q \right\rangle_W
\end{equation}
and
\begin{align}
  \label{bugger}
  D_{\delta\boldtheta} L(\q) &= ( D_{\delta\boldtheta} Q)\pd{\q}{t} + Q( D_{\delta\boldtheta} \pd{\q}{t})
                                  + \pd{}{x_i}(D_{\delta\boldtheta}(A^i \q))\nonumber\\
                             &= Q(\delta\boldtheta)\pd{\q}{t} + Q\pd{}{t}(D_{\delta\boldtheta}(\q))
                               + \pd{}{x_i}(A^i(\delta\boldtheta)\q + A^i D_{\delta\boldtheta}(\q)) \nonumber\\
                             &= 0
\end{align}
We now take the inner product of (\ref{bugger}) with a dual vector $\q^*$

\begin{equation}
  \label{knob_curd}
  \left\langle \q^*, Q(\delta\boldtheta)\pd{\q}{t} + \pd{}{x_i}(A^i(\delta\boldtheta)\q)\right\rangle_W
     + \left\langle \q^*, Q\pd{}{t}( D_{\delta\boldtheta} (\q)) + \pd{}{x_i}(A^i D_{\delta\boldtheta}(\q)) \right\rangle_W = 0
\end{equation}
We have

\begin{equation}
  Q\pd{}{t}( D_{\delta\boldtheta} (\q)) + \pd{}{x_i}(A^i D_{\delta\boldtheta}(\q)) = L(D_{\delta\boldtheta}\q)
\end{equation}
Using the definition of the adjoint map on the second term in (\ref{knob_curd}) gives

\begin{equation}
  \label{dual_id}
\left\langle
  \q^*,L_{\delta\boldtheta}(\q) \right\rangle_W +
      \left\langle L^*\q^*,D_{\delta\boldtheta}(\q)\right\rangle_W = 0
    \end{equation}
 where

\begin{equation}
      L_{\delta\boldtheta}(\q) = Q(\delta\boldtheta)\pd{\q}{t} + \pd{}{x^i}(A^i(\delta\boldtheta)\q)
\end{equation}
Adding (\ref{dual_id}) to (\ref{eq:dd_chi}) gives

\begin{align}
  D_{\delta\boldtheta}\chi &= \left\langle (\q - \boldd)\delta(x-x_r)\chi_{\mathcal{I}}, D_{\delta\boldtheta}\q \right\rangle_W
     + \left\langle \q^*,L_{\delta\boldtheta}(\q) \right\rangle_W + \left\langle L^*\q^*,D_{\delta\boldtheta}(\q)\right\rangle_W\nonumber\\
                            &= \left\langle (\q - \boldd)\delta(x-x_r)\chi_{\mathcal{I}}+ L^*\q^*, D_{\delta\boldtheta}\q \right\rangle_W
                              +  \left\langle \q^*,  L_{\delta\boldtheta}(\q)\right\rangle_W
\end{align}
$D_{\delta\boldtheta}\q$ is an expensive calculation so we define $\q^*$ to be the solution of the adjoint equation defined by

\begin{equation}
  \label{eq:adj_eq}
  L^*\q^* = - (\q - \boldd)\delta(x-x_r)\chi_\mathcal{I}
\end{equation}
with appropriate initial and boundary conditions given in the next section.
Therefore, the derivatives of the misfit functional may be calculated by

\begin{equation}
  \label{eq:adj_ker}
   D_{\delta\boldtheta}\chi = \left\langle \q^*,  L_{\delta\boldtheta}(\q)\right\rangle_W
 \end{equation}

\subsection{The formal adjoint}
We now derive the formal adjoint of $L(\q)$ with respect to the inner
product $\langle \cdot,\cdot \rangle_W$. First we note that

\begin{equation}
  \label{eq:time_ibp}
  \left\langle \q^*, Q\pd{\q}{t} \right\rangle_W =  \int_\Omega\left( \left. \bigg( Q^*\q^*, \q\bigg)_{\R^{13}}\right|_{0}^{T} - \int_0^T \left( Q^* \pd{\q^*}{t}, \q\right)_{\R^{13}} dt \right)dx
\end{equation}
where $(\cdot,\cdot)_{\R^{13}}$ is the Euclidean inner product on
$\R^{13}$ with weights $w_i$ and $Q^*$ is the adjoint of $Q$ in the
weighted inner product; in this instance, $Q^* = Q\tp$.
Typically in applications we assume that $\q(x,0)= \mathbf{0}$, while
the other boundary term vanishes if we assume
$\q^*(x,T) = \mathbf{0}$, thus the adjoint field $\q^*$ satisfies a
final value problem.  Next we deal with the spatial terms which again
are integrated by parts using Gauss' theorem.

It is convenient to write $\q$ in block form as in (\ref{eq:decomp})

$$
  \q =
  \blockvector{\q_1}{\q_2}
$$
Similarly we write $A^i$ in block form

\begin{equation}
  A^i =
\blockmatrix{0}{A_2^i}{A_1^i}{0}
\end{equation}
Then

\begin{align}
  \left\langle \q^*, \pd{}{x_i}(A^i \q) \right\rangle_W &=
              -\left\langle  \pd{}{x_i} \blockvector{\q_1^*}{\q_2^*}, \blockvector{A_2^i\q_2}{A_1^i\q_1}\right\rangle_W + \text{surface terms}\nonumber\\
                                                        &=
               -\left\langle \blockmatrix{0}{A_2^{i,*}}{A_1^{i,*}}{0}\pd{}{x_i}\blockvector{\q_1^*}{\q_2^*}, \blockvector{\q_1}{\q_2}   \right\rangle_W  + \text{surface terms}
\end{align}
where, for $i=1$,

\begin{equation}
  A_1^{1,*} = -
  \begin{pmatrix}
    1 &0 &0 &0 &0 &0 &0\\
    0 &0 &0 &1 &0 &0 &0\\
    0 &0 &0 &0 &0 &1 &0\\
    0 &0 &0 &0 &0 &0 &-1\\
    0 &0 &0 &0 &0 &0 &0\\
    0 &0 &0 &0 &0 &0 &0\\
  \end{pmatrix}
\end{equation}
and

\begin{equation}
  A_2^{1,*} = -
  \begin{pmatrix}
    2\mufr + \lambda & 0 & 0 & \alpha M &0 &0\\
    \lambda          & 0 & 0 & \alpha M &0 &0\\
    \lambda          & 0 & 0 & \alpha M &0 &0\\
    0                &\mufr &0 & 0       &0 &0\\
    0                &0   &0 & 0       &0 &0\\
    0                &0   &\mufr & 0       &0 &0\\
    -\alpha M        &0   &0 & -M       &0 &0\\
  \end{pmatrix}    
\end{equation}

We now dispose of the surface terms: we may write the boundary term as

  \begin{equation}
    \int_{\partial\Omega} q^*_j (A_{j,k}^i q_k n_i) w_j dS
  \end{equation}
We recall that the first 7 elements of $\q$ are the strain components
$(\epsilon_{11}, \epsilon_{22},\epsilon_{33}, \epsilon_{12},
\epsilon_{23}, \epsilon_{13},\zeta)\tp$. Assuming the traction-free
boundary condition in section \ref{sec:boundary-conditions} it follows
that the last 6 components of $A_{j,k}^i q_k n_i$ vanish. As we will
see below, the first 7 elements of $q^*$ may be identified as the
components of the stress tensors ${\bf \T}$ and $\Tf$, and so the
traction-free condition also implies that the first 6 components of
$q^*_j (A_{j,k}^i n_i) w_j$ vanish. In the case of absorbing boundary
conditions at artificial boundaries more care needs to be taken with
implementation to ensure that the boundary terms above vanish.

In a similar fashion we obtain

$$
A^{i,*} = \blockmatrix{0}{A_2^{i,*}}{A_1^{i,*}}{0}
$$
for $i=2,3$ where

\begin{align}
   A_1^{2,*} &= -
  \begin{pmatrix}
    0 &0 &0 &1 &0 &0 &0\\
    0 &1 &0 &0 &0 &0 &0\\
    0 &0 &0 &0 &1 &0 &0\\
    0 &0 &0 &0 &0 &0 &0\\
    0 &0 &0 &0 &0 &0 &-1\\
    0 &0 &0 &0 &0 &0 &0\\
  \end{pmatrix}
  & A_2^{2,*} = -
  \begin{pmatrix}
    0                & \lambda & 0 &0 & \alpha M &0 \\
    0                &  2\mufr + \lambda & 0 &0 & \alpha M &0 \\
    0                & \lambda & 0 &0 & \alpha M &0 \\
    \mufr                &0 &0 &0 & 0       &0 \\
    0                &0   &\mufr &0 & 0       &0 \\
    0                &0   &0 &0 & 0       &0 \\
    0                &-\alpha M        &0   &0 &-M & 0\\
  \end{pmatrix}\\
    A_1^{3,*} &= -
  \begin{pmatrix}
    0 &0 &0 &0 &0 &1 &0\\
    0 &0 &0 &0 &1 &0 &0\\
    0 &0 &1 &0 &0 &0 &0\\
    0 &0 &0 &0 &0 &0 &0\\
    0 &0 &0 &0 &0 &0 &0\\
    0 &0 &0 &0 &0 &0 &-1\\
  \end{pmatrix} 
  &A_2^{3,*} = -
  \begin{pmatrix}
    0          & 0  & \lambda &0 &0 & \alpha M \\
    0          & 0  & \lambda &0 &0 & \alpha M \\
    0          & 0  & 2\mufr + \lambda &0 &0 & \alpha M \\
    0          &0 &0  &0 &0 & 0        \\
    0          &\mufr   &0  &0 &0 & 0        \\
    \mufr      &0&0 &0 &0 &0          \\
   0           &0   & -\alpha M  &0 &0  & -M       \\
  \end{pmatrix}
\end{align}
This gives

\begin{equation}
  L^* \q^* = -Q^*\pd{\q^*}{t} - A^{i,*} \pd{\q}{x_i}
\end{equation}
Therefore, under the inner product $\langle \cdot,\cdot\rangle_W$ the
adjoint or dual map $L^*$ of $L$ is simply the non-conservative
velocity/stress formulation of the poroelastic wave equation, see
\cite{leveque} for the elastic wave case.  This permits
straightforward derivation of dual flux conditions for the adjoint
equation, as well as giving physical meaning to the adjoint.

\subsection{Dual numerical fluxes for the adjoint problem}

To derive numerical fluxes we again write $\q$ in block form

$$
\q
= \blockvector{\q_1}{\q_2}
$$
where $\q_1$ is an element of $\R^7$ and $\q_2$ of $\R^6$. The
weighted inner product on $\R^{13}$ naturally decomposes to a weighted
inner product on $\R^7$ and an unweighted inner product on $\R^6$.
Define a dual vector $\q^*$ by setting

$$
\blockvector{\q_1^*}{\q_2^*} = \blockvector{C\q_1}{\q_2}
$$
where

\begin{equation}
  C = 
  \begin{pmatrix}
    2\mufr + \lambda & \lambda &  \lambda & 0 & 0 &0 & -\alpha M \\
    \lambda &2\mufr + \lambda & \lambda  & 0 & 0 &0 & -\alpha M \\
    \lambda &\lambda &2\mufr + \lambda & 0 & 0 &0 & -\alpha M \\
    0       &0       &0       & 2\mufr & 0 &0 & 0 \\
    0       &0       &0       & 0 & 2\mufr &0 & 0\\
    0       &0        &0         & 0 & 0 &2\mufr & 0\\
   -\alpha M&-\alpha M&-\alpha M & 0 & 0 &0      &M\\
  \end{pmatrix}
\end{equation}
Note that $C^* = C$, i.e. $C$ is self-adjoint in the weighted inner
product on $\R^7$.
Let $D^k$ be an element, then (recalling equation \eqref{eq:mass_matrix}) we have

\begin{multline}
  \int_{D^k} \left(Q_1 \pd{\q_1^k}{t} + \pd{}{x_i}(A^i_2 \q_2^k), C\p_1 \right)_{\R^{7}}
  + \left( Q_2 \pd{\q_2^k}{t} + \pd{}{x_i}(A^i_1 \q_1^k), \p_2 \right)_{\R^{6}}dx \\
     = \int_{D^k} \left(Q_1 \pd{(C \q_1^k)}{t} + C\pd{}{x_i}(A^i_2 \q_2^k), \p_1 \right)_{\R^{7}}
  + \left( Q_2 \pd{\q_2^k}{t} + \pd{}{x_i}(A^i_1 \q_1^k), \p_2 \right)_{\R^{6}}dx 
\end{multline}
Using $CA^i_2=A_2^{i,*}$ and $A_1^i = A_1^{i,*}C$, the following identities
are easily derived:

\begin{align}
  C \pd{}{x_i}(A^i_2 \q_2) &= A_2^{i,*} \pd{\q_2^*}{x_i}\\
  \pd{}{x_i}(A_1^i\q_1)      &= A_1^{i,*} \pd{\q_1^*}{x_i}
\end{align}
where for notational convenience we have suppressed the dependency on the element $D^k$.
This gives

\begin{multline}
  \int_{D^k} \left(Q_1 \pd{\q_1^k}{t} + \pd{}{x_i}(A^i_2 \q_2^k), C\p_1 \right)_{\R^{7}}
  + \left( Q_2 \pd{\q_2^k}{t} + \pd{}{x_i}(A^i_1 \q_1^k), \p_2 \right)_{\R^{6}}dx \\
     = \int_{D^k} \left(Q_1 \pd{\q_1^*}{t} + A_2^{i,*} \pd{\q_2^*}{x_i}, \p_1 \right)_{\R^{7}}
  + \left( Q_2 \pd{\q_2^*}{t} + A_1^{i,*} \pd{\q_1^*}{x_i}, \p_2 \right)_{\R^{6}}dx 
\end{multline}
This means that a numerical scheme for the forward model automatically
gives a scheme for the adjoint model by setting the fluxes as follows:

\begin{equation}
  \mathcal{F}^* \blockvector{\q_1^*}{\q_2^*} = \mathcal{F}^*\blockvector{C\q_1}{\q_2}
\end{equation}
That is we simply replace $\q_1$ by $C\q_1$ in the flux terms for the
forward model in section (\ref{sec:flux_poro}).

We obtain the following upwind flux:

\begin{multline}
  \beta_1 (\cpI)^- \rr_1^- + \beta_2 c_s^- \rr_2^- + \beta_3 c_s^- \rr_3^- + \beta_3 (\cpII)^- \rr_4^-= \\
  (\cpI)^-(d_{11} \nt\sbb{\T} + d_{12} \nt\sbb{\Tf} + d_{13}\sbb{\vvs} + d_{14}\sbb{\vvf}) \times
  \begin{pmatrix}
    \overline{2\mufr^- \n\otimes\n + (\lambda^- + \alpha^- \gamma_1^- M^- )I}\\
    -(\alpha^-  + \gamma_1^-) M^-\\
    (\cpI)^-\n\\
    \gamma_1^- (\cpI)^-\n
  \end{pmatrix}\\
   - \frac{(c_s)^-(c_s)^+}{ (c_s)^+(\mufr)^- + (c_s)^-(\mufr)^+ }
   \begin{pmatrix}
     \overline{2\mufr^- \sym(\n\otimes(\n\times(\n\times\sbb{\T})))}\\
     0\\
     (c_s)^- \n\times(\n\times\sbb{\T})\\
     \frac{-(c_s)^- \rho_f^-}{m^-} \n\times(\n\times\sbb{\T})
   \end{pmatrix} \\ %
   - \frac{(c_s)^-(\mufr)^+}{ (c_s)^+(\mufr)^- + (c_s)^-(\mufr)^+ }
   \begin{pmatrix}
     \overline{2\mufr^-\sym(\n\otimes(\n\times(\n\times\sbb{\vvs})))}\\
     0\\
     (c_s)^- \n\times(\n\times\ssb{\vvs})\\
     \frac{-(c_s)^- \rho_f^-}{m^-} \n\times(\n\times\ssb{\vvs})
   \end{pmatrix}\\
   + (\cpII)^-(d_{21} \nt\sbb{\T} + d_{22} \nt\sbb{\Tf} + d_{23}\sbb{\vvs} + d_{24}\sbb{\vvf}) \times\\
   \begin{pmatrix}
     \overline{2\mufr^-\n\otimes\n + (\lambda^- +  \alpha^- \gamma_2 ^- M^- )I }\\
     -(\alpha^-  + \gamma_2^-) M^-\\ 
     (\cpII)^-\n\\
     \gamma_2^- (\cpII)^-\n
   \end{pmatrix}
\end{multline}

\subsection{Dual numerical fluxes for coupled elastic/poroelastic problems}

In a similar manner one can derive dual numerical fluxes for elastic
and coupled elastic and poroelastic problems, which we state below for
convenience. The elastic case can be found in \cite{wilcox10} and is
repeated here for completeness and notational consistency.

For the elastic case we have:

\begin{multline}
  \beta_1 (c_p)^- \rr_1^- + \beta_2 c_s^- \rr_2^- + \beta_3 c_s^- \rr_3^- = \\
  \frac{(c_p)^- c_p^+ \nt\sbb{\boldS} + (c_p)^- (\lambdae^+ + 2\mue^+)\sbb{\vve}}
       {c_p^+(\lambdae^-+2\mue^-) + c_p^-(\lambdae^++2\mue^+)}
       \begin{pmatrix}
        \overline{2 \mue^- \n\otimes\n + \lambda_e^- I}\\
         (c_p)^- \n
       \end{pmatrix}\\
       - \frac{(c_s)^-(c_s)^+}{ (c_s)^+(\mue)^- + (c_s)^-(\mue)^+ }
       \begin{pmatrix}
         \overline{2 \mue^- \sym(\n\otimes(\n\times(\n\times\sbb{\T})))}\\
         (c_s)^-\n\times(\n\times\sbb{\T})
       \end{pmatrix}\\
       - \frac{(c_s)^-(\mue)^+}{ (c_s)^+(\mue)^- + (c_s)^-(\mue)^+ }
       \begin{pmatrix}
         \overline{2 \mue^- \sym(\n\otimes(\n\times(\n\times\sbb{\vve})))}\\
         (c_s)^- \n\times(\n\times\ssb{\vvs})
       \end{pmatrix}
\end{multline}

For an interface element on the elastic domain  we define an upwind flux by:

\begin{multline} \beta_1^e (c_p^e)^- \rr_1^{-,e} + \beta_2 c_s^- \rr_2^- + \beta_3 c_s^- \rr_3^- = \\
  (c_p^e)^-(d_{11} \nt\sbb{\boldS,\T} + d_{12} \sbb{\vve,\vvs} + d_{13}\sbb{\vvf} ) \times
  \begin{pmatrix}
    \overline{2\mue^- \n\otimes\n + \lambdae^- I}\\
    (c_p)^- \n
  \end{pmatrix} \\
  - \frac{(c_s^e)^-(c_s)^+}{ (c_s)^+(\mue)^- + (c_s^e)^-(\mufr)^+ }
  \begin{pmatrix}
    \overline{2\mue^- \sym(\n\otimes(\n\times(\n\times\sbb{\boldS, \T})))}\\
    (c_s)^- \n\times(\n\times\sbb{\boldS,\T})
  \end{pmatrix}\\
  - \frac{(c_s^e)^-(\mufr)^+}{ (c_s)^+(\mue)^- + (c_s)^-(\mufr)^+ }
  \begin{pmatrix}
    \overline{2\mue^-\sym(\n\otimes(\n\times(\n\times\ssb{\vve, \vvs})))}\\
    (c_s)^- \n\times(\n\times\ssb{\vve,\vvs})
  \end{pmatrix}
\end{multline}

Finally, for an interface element on the poroelastic domain we have

\begin{multline}
  \beta_{10}^p (\cpII)^+ \rr_{10}^+ + \beta_{11}^p (c_s^p)^+ \rr_{11}^+ + \beta_{12}^p (c_s^p)^+ \rr_{12}^+ + \beta_{13}^p (\cpI)^+ \rr_{13}^+ = \\
  (\cpII)^+(\tilde{d}_{21} \nt\sbb{\boldS,\T} + \tilde{d}_{22} \sbb{\vve,\vvs} + \tilde{d}_{23}\sbb{\vvf} ) \times
  \begin{pmatrix}
    \overline{2\mufr^+ \n\otimes\n + (\lambda^+ + \alpha^+ \gamma_2^+ M^+ )I }\\
    -(\alpha^+ + \gamma_2^+)M^+\\
    -(\cpII)^+\n\\
    -\gamma_2^+ (\cpII)^+\n
  \end{pmatrix}\\
  - \frac{1}{(c_s^p)^+(\mue)^- + (c_s^e)^-(\mufr)^+}\times
     \left\{ (c_s^e)^-(c_s^p)^+
       \begin{pmatrix}
         \overline{2\mufr^+ \sym(\n\otimes(\n\times(\n\times\sbb{\boldS,\T})))}\\
         0\\
         -(c_s)^+ \n\times(\n\times\sbb{\boldS,\T})\\
         \frac{(c_s)^+\rho_f^+}{m^+} \n\times(\n\times\sbb{\boldS,\T})
       \end{pmatrix} \right .  \\ %
     \left .  - \mue^- (c_s^p)^+
       \begin{pmatrix}
         \overline{2\mufr^+ \sym(\n\otimes(\n\times(\n\times\ssb{\vve,\vvs})))}\\
         0\\
         -(c_s)^+ \n\times(\n\times\ssb{\vve,\vvs})\\
         \frac{(c_s)^+ \rho_f^+}{m^+} \n\times(\n\times\ssb{\vve,\vvs})
       \end{pmatrix}
     \right\}\\
     + (\cpI)^+(\tilde{d}_{31}\nt\sbb{\boldS,\T} + \tilde{d}_{32} \sbb{\vve,\vvs} + \tilde{d}_{33}\sbb{\vvf} ) \times%
     \begin{pmatrix}
        \overline{2\mufr^+ \n\otimes\n  + (\lambda^+ + \alpha^+ \gamma_1^+ M^+ )I} \\
       -(\alpha^+ + \gamma_1^+)M^+\\ 
       -(\cpI)^+\n\\
       -\gamma_1^+ (\cpI)^+\n
     \end{pmatrix}
\end{multline}

\subsection{Discussion}

\subsubsection{Implementation} It turns out that implementation of the
adjoint method to estimating derivatives of an objective functional is
quite straightforward as we now show. Once again it is convenient to
write $\q$ and $\q^*$ in block form:

$$
\q
= \blockvector{\q_1}{\q_2}, \quad
\q^*
= \blockvector{\q_1^*}{\q_2^*}
$$
Then

\begin{align}
  L_{\delta \theta} (\q) &= \blockmatrix{I}{0}{0}{Q_2(\delta\theta)} \pd{}{t} \blockvector{\q_1}{\q_2} + \pd{}{x} \blockmatrix{0}{A_2^i}{A_1^i(\delta\theta)}{0} \blockvector{\q_1}{\q_2}\\
                     &= Q_2(\delta\theta) \pd{\q_2}{t} + \pd{}{x}(A_1^i(\delta\theta)\q_1)
\end{align}
since from equation \eqref{hyp_sys_mat} we have

$$
\pd{\q_1}{t} + \pd{}{x}(A_2^i\q_2) = 0
$$
This means that (\ref{eq:adj_ker}) reduces to computing

\begin{equation}
  \label{eq:adj_ker_simp}
  D_{\delta\boldtheta}\chi = \left\langle \q_2^*,  L_{\delta\boldtheta}^2(\q)\right\rangle_{W_2}
\end{equation}
where

$$
  L_{\delta\boldtheta}^2(\q) = Q_2(\delta\theta) \pd{\q_2}{t} + \pd{}{x}(A_2^i(\delta\theta)\q_1)
$$
and $W_2$ is the restriction of $W$ to $\q_2$, the velocity components
of $\q$.  This means that to compute $D_{\delta\boldtheta}\chi$ we
only need the velocity components $\q_2$ and $\q_2^*$ of $\q$ and
$\q^*$.  Thus for implementation it is immaterial whether we use a
conservative velocity/strain or non-conservative velocity/stress
(adjoint) formulation to compute $\q^*$ since we only need $\q^*_2$.

\subsubsection{Time reversal} Implementation of a time-reversed
adjoint solver needs some care since the downwind fluxes given above
are with respect to forward time integration. Integrating backwards
from the final time $t=T$ to $t=0$ they become upwind fluxes and
result in a divergent scheme. To obtain a downwind scheme one simply
has to map the wavespeeds $c \rightarrow -c$.

\subsubsection{Fr\'echet kernels of poroelastic parameters}
Sensitivity or Fr\'echet kernels obtained from (\ref{eq:adj_ker}) by
taking the integral with respect to time are a useful tool in
computational seismology; we refer to \cite{fichtner11}, Chapter 9,
and \cite{tromp08} for the elastic case. Due to the nonlinear
relationships between the constitutive parameters in the Hooke's laws
(\ref{hooke1})-(\ref{solid_tensor}) and the physical parameters in
equation (\ref{bigB})-(\ref{bigM}), Fr\'echet kernels corresponding to
the primary physical constants like porosity would be unwieldy.
Therefore, in the following, we use the derived model parameters
$\rhoa, \rhof$ and $m$ for densities, $\kappafr$ and $\mufr$ for
stiffness parameters and $\alpha$ and $M$ for coupling parameters.

For the density parameters we obtain kernels $k_{\rhoa}$,
$k_{\rhof}$ and $k_m$ given by

\begin{align}
  k_{\rhoa} &= \int_0^T \left(\uu^*, \pd{\uu}{t}\right)_{\R^3} dt\\
  k_{\rhof} &= \int_0^T \left(\uu^*, \pd{\uuf}{t}\right)_{\R^3} + \left(\uuf^*, \pd{\uu}{t}\right)_{\R^3} dt\\
  k_m      &= \int_0^T \left(\uuf^*, \pd{\uuf}{t}\right)_{\R^3} dt\
\end{align}
For the stiffness parameters we obtain kernels
$k_{\kappafr}$ and $k_{\mufr}$ where

\begin{align}
  k_{\kappafr} &= -\int_0^T \left(\uu^*, \nabla \trace(E)\right)_{\R^3} dt\\
  k_{\mufr}   &= -\int_0^T \left(\uu^*, \nabla\cdot E - \frac{1}{3} \nabla \trace(E)\right)_{\R^3} dt
\end{align}
For the coupling coefficients we first define an auxiliary kernel
$k_{\alpha,M}$ by

\begin{equation}
  k_{\alpha,M} = \int_0^T \left(\uu,\nabla \zeta \right)_{\R^3} - \left(\uuf,\nabla\trace(E) \right)_{\R^3} dt
\end{equation}
This gives kernels $k_\alpha$ and $k_M$ defined by

\begin{align}
  k_\alpha &= Mk_{\alpha,M}\\
  k_M     &= \alpha k_{\alpha,M} + \int_0^T \left(\uuf,\nabla \zeta \right)_{\R^3}dt
\end{align}

We may then write

\begin{equation}
  D_{\delta\boldtheta}\chi = \int_\Omega (\delta\rhoa) k_{\rhoa} + (\delta\rhof) k_{\rhof}
  + (\delta m) k_{m}  + (\delta\kappafr) k_{\kappafr} + (\delta\mufr) k_{\mufr}
      + (\delta\alpha) k_{\alpha} + (\delta M) k_{M} dx
\end{equation}

\section{Numerical experiments}
\label{sec:numer-exper}

In this section, we consider several numerical experiments.  First, we
consider the convergence properties of the numerical scheme in the
inviscid and low- and high-frequency viscous regimes; we verify that,
except in some cases of very small permeability, our code approaches
the optimal convergence behaviour of the DG method (see discussion in
\cite[Chapter 4]{hesthaven_warburton_book} and references therein).
We then give an example of heterogeneous poroelastic material to
show that our code naturally handles material discontinuities, a
necessary feature in applications to groundwater tomography. Finally
we give an example of the adjoint method.

In the simulations described below, the length of the time step $\Delta t$
is computed from
\begin{equation}
  \label{eq:CFL}
  \Delta t = C\frac{h_{\min}}{c_{\max}p^2}
\end{equation}
where $C$ is a constant, $c_{\max}$ is the maximum wave speed over all
elements, $p$ is the basis order and $h_{\min}$ is the smallest
distance between two vertices in any element. In the simulations, we
set $C=0.4$ unless otherwise stated.

\subsection{Convergence analysis}
\label{sec: conv_anal}

Convergence tests were carried out on a cubical domain
$\Omega=[0,5]\times[0,5]\times[0,5]$\,m with four regular grids of
different side lengths (formed by dividing the domain into subcubes
and dividing each subcube into tetrahedra) and inhomogeneous Dirichlet
boundary conditions.  For time-stepping, in this section we used the
five-stage, fourth-order accurate low-storage explicit Runge-Kutta
(LSERK) method originated in \cite{carpenter94} and used in
\cite{hesthaven_warburton_book}. With three-dimensional meshes, the
advantages of low-storage methods, storing fewer intermediate results
than general Runge-Kutta methods, become particularly apparent.

The material parameters are given in Table \ref{tab:params_conve}. We consider
three cases. In the first case we consider wave propagation in an
inviscid setting, while the other two involve viscous flow in Biot's
low- and high-frequency settings respectively. In Table
\ref{tab:derived}, we list the assumed frequencies, viscosities,
permeabilities, and the derived wave velocities. The frequency was set
at 2,000 Hz so that the test domain captured around three wavelengths
of the fast P-wave. Note that with the high-frequency case we also
need to define the quality factor (see Section~\ref{high_freq}).

Analytic plane wave solutions consisting of fast and slow P-waves and
S-waves were constructed from plane wave solutions of the form

$$
\q = \q_0 e^{\i(k_x x + k_y y +k_z z - \omega t)}
$$
where $\i=\sqrt{-1}$, $\omega$ is an angular frequency, and $k_x$, $k_y$ and $k_z$ are
complex wave numbers in the $x$-, $y$- and $z$-directions, respectively. In
the inviscid case, we consider dissipating waves of the form
$$
\q = \Re\left( \sum_{j=1}^{13} \alpha_j \rr_j e^{\i(k_{x,j}  + k_{y,j} +k_{z,j} - \omega t)}\right)
$$
where $\rr_j$ is an eigenvector of the $13\times 13$ matrix
$$
\Pi = Q^{-1}(\hat{n}_x A^1 + \hat{n}_y A^2 + \hat{n}_z A^3)
$$
where $n_x$, $n_y$ and $n_z$ are direction cosines. In the reported
examples, we set $[k_x, k_y, k_z]$ to be a vector parallel to
$[0.9, 1.0, 1.1]$, so as not to align with the geometry of the regular
grid in use. For the viscous low- and high-frequency cases the wave
speeds and dissipation are frequency-dependent.

\begin{table}[!ht]
  \centering
  \caption{Material parameters used in the convergence analysis.}\label{tab:params_conve}
  \begin{tabular}{c|cc}
    \hline
    \textbf{variable name} & \textbf{symbol}  & \\
    \hline
    solid density &  $\rhos$ (kg/m$^{3}$)  & 2650\\
    fluid density & $\rhof$ (kg/m$^{3}$)  & 900\\
    fluid bulk modulus &  $\kappaf$ (GPa)  &  2.0\\
    frame bulk modulus &  $\kappafr$ (GPa)  &  10.0\\
    solid bulk modulus &  $\kappas$ (GPa)   &  12.0\\
    frame shear modulus &  $\mufr$ (GPa)  &    5.0\\
    tortuosity &  $\tau$ &  1.2\\
    porosity &  $\phi$  &  0.3\\
    \hline
  \end{tabular}
\end{table}

\begin{table}[!h]
  \caption{This table lists the plane wave frequency $f_0$, viscosity $\eta$, 
    permeability $k$, quality factor $Q_0$, Biot's characteristic frequency $f_c$, 
    and wave velocities $(\cpI, \cpII, c_s)$ for the three
    cases studied.}\label{tab:derived} 
  \begin{center}
    \begin{small}
      \begin{tabular}{ccccc||cccc}
        \hline
        case  & $f_0$ (Hz) & $\eta$ (Pa$\cdot$s) & $k$ (m$^2$) & $Q_0$ & $f_c$ (Hz)& $\cpI$ (m/s)& $\cpII$ (m/s)& $c_s$ (m/s)\\
        \hline
       inviscid        & 2000  & 0      & -         & -  & -  &2967 &1411 &1622 \\
        low-frequency   & 2000  & 0.001  & 10$^{-12}$ & -  & 44209.71 &2817 &414& 1534\\
        high-frequency  & 2000  & 0.001  & 10$^{-8}$  &30  & 4.42 &2967& 1411 &1622\\
        \hline
      \end{tabular}
    \end{small}
  \end{center}
\end{table}

The numerical solver was initialised with the analytic plane wave
solution at time $t=0$, and the boundary values were set with the
values of the analytic plane wave.  The tests were carried out using
plane waves with a fixed frequency $f_0$ (see Table
\ref{tab:derived}). The total simulation time was taken to be $1/f_0$.
The analytic and numerical solutions were compared at the final
simulation time over the whole computational domain $\Omega$ by, on
each element $D^k$, interpolating a polynomial of degree at most $p$
through the exact solution values, calculating the distance in
$L^2(D^k)$ between this polynomial and the polynomial representing the
simulated solution, and combining the results over all elements to
give a distance in $L^2(\Omega)$. Errors are reported only for the
solid velocity component $\us$ in all cases.

The convergence rate is defined by
\begin{equation}
  \label{eq:convorder}
  \mbox{rate} = \log{\left(\frac{\|e^{\ell}\|_2}{\|e^{\ell-1}\|_2}\right)}\bigg/\log{\left(\frac{h_{\min}^{\ell}}{h_{\min}^{\ell-1}}\right)}
\end{equation}
where $\|e^{\ell}\|_2$ is the $L^2$ norm of the error $e^{\ell}$ as
described above and $h_{\min}^{\ell}$ is the minimal distance between
adjacent vertices in the $\ell$'th mesh; here the meshes are ordered
in decreasing order of $h_{\min}$.

Table \ref{tab:order} shows the convergence rate for the inviscid,
viscous (low-frequency), and viscous (high-frequency) cases.  The
results shows that the method is consistent with the optimal $p+1$, 
for order $p$.

\begin{table*}[!h]
  \caption{The convergence rate as a function of the grid parameter
    $h_{\min}$ for three basis orders starting from order 3 (top),
    order 5 (middle), and ending with order 6 (bottom). Convergence
    rates together with the $L^2$-error values are reported for the
    inviscid (columns 2 and 3), viscous (low-frequency, columns 4 and
    5), and viscous (high-frequency, columns 6 and 7) cases.}\label{tab:order}
  \begin{center}
    \begin{small}
      \begin{tabular}{c|cc|cc|cc}
        \hline
        &\multicolumn{2}{|c}{{\bf inviscid}}&\multicolumn{2}{|c}{{\bf
            low-frequency}}&\multicolumn{2}{|c}{{\bf high-frequency}}\\
        $h_{\min}$ (m) & $L^2$ error   & rate& $L^2$ error   & rate& $L^2$ error   & rate\\
        \hline
0.3125 & 2.032e-01 & ---    & 2.020e-01 & ---    & 1.986e-01 & ---    \\
0.2632 & 1.051e-01 & 3.8362 & 1.046e-01 & 3.8301 & 1.031e-01 & 3.8157 \\
0.2083 & 4.059e-02 & 4.0728 & 4.042e-02 & 4.0695 & 3.977e-02 & 4.0780 \\
0.1786 & 2.186e-02 & 4.0140 & 2.171e-02 & 4.0323 & 2.144e-02 & 4.0088 \\\hline
0.3125 & 6.640e-03 & ---    & 7.196e-03 & ---    & 6.476e-03 & ---    \\
0.2632 & 2.432e-03 & 5.8448 & 2.808e-03 & 5.4762 & 2.382e-03 & 5.8206 \\
0.2083 & 6.415e-04 & 5.7045 & 7.130e-04 & 5.8679 & 6.283e-04 & 5.7043 \\
0.1786 & 2.446e-04 & 6.2553 & 2.594e-04 & 6.5585 & 2.397e-04 & 6.2500 \\\hline
0.3125 & 1.033e-03 & ---    & 1.125e-03 & ---    & 1.005e-03 & ---    \\
0.2632 & 3.233e-04 & 6.7586 & 3.835e-04 & 6.2620 & 3.151e-04 & 6.7513 \\
0.2083 & 6.375e-05 & 6.9503 & 8.913e-05 & 6.2462 & 6.237e-05 & 6.9334 \\
0.1786 & 2.165e-05 & 7.0072 & 3.547e-05 & 5.9780 & 2.122e-05 & 6.9944 \\\hline
        \hline
    \end{tabular}
    \end{small}
  \end{center}
\end{table*}

\subsubsection{The low frequency case: very small permeability}\label{sec:stiff}

As noted in the introduction to Section~\ref{sec:dissip}, the accuracy
of the low-storage explicit Runge-Kutta (LSERK) scheme falls off as
the permeability decreases to zero in the low frequency regime. In
this section we give convergence results for an example in which the
permeability is $k=10^{-14}$ m${}^2$, which may be regarded as a
fairly extreme test of a time integration scheme. On the meshes and
basis orders used for Table~\ref{tab:order}, the LSERK scheme failed
in every case, with all fields rapidly diverging to $\infty$.

For small $k$, stiffness is introduced into the system by the
low-frequency dissipation terms $\g$ described by
(\ref{low_freq_source}); having no space derivatives, these play no
role in the spatial semidiscretisation of the system, and appear
unchanged in the ODE system, where they are localised on individual
nodes. To deal with them, we tried two techniques: operator splitting
and an IMEX (implicit-explicit) Runge-Kutta scheme. In both
approaches, the idea is to regard the right-hand side of the ODE
system as the sum of two terms: the stiff term $\g$ and the remaining,
non-stiff, conservation terms.

In a recent paper \cite{Shukla2018} on the Biot equation in two
dimensions, it is observed that the ODE system with only the stiff
terms on the right-hand side may be solved explicitly. This remains
true in three dimensions: if we compute $Q^{-1}\g$, we find a matrix
that is zero outside the lower-right $6\times 6$ block; this block,
itself broken down into $3\times 3$ blocks, acts on the velocity terms
as follows
\begin{equation}\label{eq:stiff_rhs}
\begin{bmatrix}
\rhoa I & \rhof I \\
\rhof I & mI
\end{bmatrix}^{-1}
\begin{bmatrix}
0 & 0 \\
0 & -(\eta/k) I
\end{bmatrix}
\begin{bmatrix}
* \\ * \\ * \\ \uf \\ \vf \\ \wf
\end{bmatrix}
=
\frac{\eta}{(m\rhoa-\rhof^2)k}
\begin{bmatrix}
0 & \rhof I \\
0 & \rhoa I
\end{bmatrix}
\begin{bmatrix}
* \\ * \\ * \\ \uf \\ \vf \\ \wf
\end{bmatrix}
\end{equation}
Here $0$, $I$ represent the $3\times3$ zero and identity matrices and
asterisks denote terms that are multiplied by zero, so have no part to
play.

Diagonalising this triangular matrix is entirely straightforward: its
eigenvalues are $0$ and $\rhoa\eta/((m\rhoa-\rhof^2)k)$ and its
eigenvectors are readily obtained, leading to a simple, explicit
solution to the associated ODE system.

We can now follow \cite{Shukla2018} and implement Godunov
splitting \cite[Section 17.3]{leveque}: at each time-step, given an 
initial value $\q_n$ at time $t_n$ from the previous timestep, we begin by
explicitly finding the solution to the stiff part of the system at the next
time-step, $t_{n+1}$; we then feed this back as a new initial value at 
$t_n$ and from that use the LSERK scheme to find an approximate solution to
the non-stiff part of the system at $t_{n+1}$. This serves as our
approximate solution $\q_{n+1}$ of the whole system at $t_{n+1}$, and
we can repeat the process.

This immediately results in a stable scheme, but the errors involved
in this splitting method are rather large: first-order in the length
of the time step \cite[Section 17.3]{leveque}. In an attempt to
mitigate this, we also considered Strang splitting 
\cite[Section 17.4]{leveque}: instead of
a whole time-step of the analytic stiff solution followed by a whole
timestep of the LSERK non-stiff solution, this comprises half a time
step of analytic stiff, a whole timestep of LSERK non-stiff, and a
final half time step of analytic stiff. As in the Godunov splitting,
the final values of the system at the end of each (partial) time-step
are fed back as initial values to the next (partial) time-step. This
has scarcely any more computational cost (compared to an LSERK step,
the cost of the analytic solution is vanishingly small), and should
improve the time-stepping error to second-order accuracy \cite[Section
17.4]{leveque}.

Table~\ref{tab:order-splitting} shows the errors and convergence rates
for a few examples, using time-steps $\Delta t$ and $\Delta t/16$
(intermediate $\Delta t/2^n$ results were calculated but are not
presented here). As expected, Strang splitting gives better results
than Godunov splitting (although the difference is not huge; it is
noted in \cite[Section 17.5]{leveque} that this is not uncommon). Both
methods give noticeably better results when the time-step length is
decreased; this is in marked contrast to the non-stiff results in
Table~\ref{tab:order}, which remain unchanged to four or more decimal
places when the time-step is halved. This suggests that, in
Table~\ref{tab:order}, we are seeing almost entirely spatial
discretisation errors, with little contribution from time
discretisation, whereas in Table~\ref{tab:order-splitting}, time
discretisation is still making a noticeable contribution to the error,
even at 16 times the base number of steps. At order $3$, we can
approach the optimal convergence rate of $4$, but only by
significantly reducing the time-step. At order $5$, even reducing the
time-step by a factor of $16$ does not give anything close to the
optimal rate, but even so we do see the errors being greatly
reduced. In summary, the splitting methods are an effective, but
possibly sub-optimal and certainly costly, way of addressing the
stiffness caused by very small permeability.

\begin{table}
\caption{Convergence rates for the stiff case, using Godunov splitting (top) and Strang splitting (bottom), with three meshes and time steps $\Delta t$ and $\Delta t/16$. Again, only the solid velocity component $\us$ is reported. Some values are highlighted for comparison with Table \ref{tab:order-IMEX}.}
\label{tab:order-splitting}
\begin{center}
\begin{tabular}{ccccccccc}
&\multicolumn{4}{c}{Order 3} & \multicolumn{4}{c}{Order 5} \\
&\multicolumn{2}{c}{Time step / 1} & \multicolumn{2}{c}{Time step / 16} & \multicolumn{2}{c}{Time step / 1} & \multicolumn{2}{c}{Time step / 16} \\
$h_{\min}$ & $L^2$ error & rate & $L^2$ error & rate & $L^2$ error & rate & $L^2$ error & rate \\\hline
0.3125 & 2.062e-01 & ---    & 1.779e-01 & ---    & 2.703e-02 & ---    & \textbf{7.247e-03} & ---    \\
0.2632 & 1.192e-01 & 3.1881 & 9.289e-02 & 3.7802 & 2.200e-02 & 1.1981 & \textbf{3.491e-03} & \textbf{4.2502} \\
0.2083 & 6.359e-02 & 2.6914 & 3.590e-02 & 4.0694 & 1.709e-02 & 1.0801 & \textbf{1.640e-03} & \textbf{3.2347} \\\hline
0.3125 & 2.020e-01 & ---    & 1.778e-01 & ---    & 2.340e-02 & ---    & \textbf{7.256e-03} & ---    \\
0.2632 & 1.147e-01 & 3.2903 & 9.286e-02 & 3.7794 & 1.879e-02 & 1.2752 & \textbf{3.488e-03} & \textbf{4.2625} \\
0.2083 & 5.839e-02 & 2.8920 & 3.589e-02 & 4.0696 & 1.448e-02 & 1.1165 & \textbf{1.611e-03} & \textbf{3.3057} \\\hline
\end{tabular}
\end{center}
\end{table}

For a less costly solution, we turned to an IMEX (implicit-explicit)
Runge-Kutta scheme. As for the explicit scheme, the size of the meshes
involved in three-dimensional simulation makes a low-storage scheme
very attractive. Several such schemes are presented in
\cite{CavaglieriBewley2015}; we used the four-stage, third-order
accurate scheme IMEXRKCB3e \cite[equation
(30)]{CavaglieriBewley2015}. In an IMEX scheme, the ODE is split as
above into a non-stiff and a stiff part; at each stage of each
Runge-Kutta step, the non-stiff part of the equation is handled
explicitly (i.e. by evaluating the non-stiff part of the right-hand
side) and the stiff part is handled implicitly (i.e. by solving an
equation involving the stiff part of the right-hand side). This
equation-solving process can, in general, be computationally
expensive, but for the low-frequency terms in Biot's equation this
turns out not to be the case. The main reason for this is that the
dissipation terms are localised onto individual nodes; this
immediately means that the equations to be solved decouple into at
worst one $13\times13$ linear system for each node.  In fact, they are
much simpler than that. As above, the dissipation terms involve only
the last six of the thirteen fields in the model, so we only need a
$6\times 6$ system. At each Runge-Kutta stage, we must \cite[Section
1.2.1]{CavaglieriBewley2015}, for each node, solve one linear system
by finding $(I-cA)^{-1}A$, where $c$ is some scalar depending on the
IMEX coefficients and the time-step length and $A$ is the matrix given
above in (\ref{eq:stiff_rhs}). The simple structure of this matrix
leads to a simple solution: in block form,
$$(I-cA)^{-1}A=
\frac{\eta}{c\eta\rhoa+km\rhoa-k\rhof^2}
\begin{bmatrix}
0 &  \rhof I \\
0 & -\rhoa I
\end{bmatrix}
$$
where $0$, $I$ again represent the $3\times3$ zero and identity
matrices.  For this system, then, the implicit part of the IMEX scheme
becomes fully explicit and the cost of the IMEX scheme is little more
than that of an LSERK scheme of the same accuracy. In fact, we used a
four-stage scheme with third-order accuracy, which is adequate for
these tests (this was verified by re-running tests with half the
time-step, which led to changes only in the fourth or more significant
figure of the error).

The results of this, on the same meshes as were used for
Table~\ref{tab:order}, are shown in Table~\ref{tab:order-IMEX}. As can
be seen, the convergence rates are consistent with the optimal rate of
$p+1$ at basis order $p$ for $p=2$ and $p=3$, marginal at $p=4$ and
fall away for $p=5$ and $p=6$. Unlike in the operator-splitting
methods, halving the time-step had no noticeable effect on this (the
results typically agreed to three or more significant figures), so
this seems to be a feature of the spatial discretisation, not of the
time-stepping. This is also consistent with the way that, in the
operator-splitting approach (Table~\ref{tab:order-splitting}), the
optimal convergence rate is apparent at basis order $p=3$ but not at
$p=5$.

\begin{table}
\caption{Convergence rates for the stiff case, IMEX scheme IMEXRKCB3e for basis orders 2--6 (columns) and four meshes (rows). Again, only the solid velocity component $\us$ is reported. Some values are highlighted for comparison with Table \ref{tab:order-splitting}.}
\label{tab:order-IMEX}
\begin{center}
\begin{tabular}{ccccccccccc}
&\multicolumn{2}{c}{Order 2} & \multicolumn{2}{c}{Order 3} & \multicolumn{2}{c}{Order 4} \\
$h_{\min}$ & $L^2$ error & rate & $L^2$ error & rate & $L^2$ error & rate \\\hline
0.3125 & 7.010e-01 & ---    & 1.770e-01 & ---    & 3.092e-02 & ---     \\
0.2632 & 5.509e-01 & 1.4025 & 9.236e-02 & 3.7851 & 1.681e-02 & 3.5484  \\
0.2083 & 2.450e-01 & 3.4676 & 3.557e-02 & 4.0840 & 4.988e-03 & 5.1996  \\
0.1786 & 1.526e-01 & 3.0709 & 1.917e-02 & 4.0107 & 2.616e-03 & 4.1875  \\
& \multicolumn{2}{c}{Order 5} & \multicolumn{2}{c}{Order 6} \\
$h_{\min}$ & $L^2$ error & rate & $L^2$ error & rate \\\hline
0.3125 & \textbf{7.194e-03} & ---    & 2.907e-03 & ---    \\
0.2632 & \textbf{3.428e-03} & \textbf{4.3137} & 1.948e-03 & 2.3300 \\
0.2083 & \textbf{1.562e-03} & \textbf{3.3632} & 1.155e-03 & 2.2363 \\
0.1786 & 1.052e-03 & 2.5655 & 7.939e-04 & 2.4342 \\
\end{tabular}
\end{center}
\end{table}

Comparing the results for operator-splitting and IMEX, we can see that
the IMEX method easily out-performs operator-splitting. As an
illustration, at basis order 5, the IMEX results are closely
comparable to the operator-splitting results (shown in bold in
Tables~\ref{tab:order-splitting} and~\ref{tab:order-IMEX}), but only
with the time-step for the operator-splitting reduced by a factor of
16.

The loss of the optimal convergence rate for larger basis orders is of
some concern. It should be noted, though, that the optimal convergence
rate is derived (e.g. \cite[\S4.5]{hesthaven_warburton_book} in one
space dimension) without source terms; a suggestion, for this rather
extreme value of permeability, is that, as the basis order $p$
increases, the error associated with the dissipation terms $\g$
decreases more slowly than that associated with the conservation part
of the equation, and at around $p=4$ or $p=5$ becomes dominant. From
that point on, the rates are largely determined by the behaviour of
$\g$, and we have no reason to expect a rate of $p+1$. Looking back at
the non-stiff case in Table~\ref{tab:order}, we can perhaps see the
beginnings of this phenomenon: at order 3, there is little difference
between the inviscid, low-frequency and high-frequency regimes, but at
order 6 the low-frequency rates are noticeably, although not greatly,
smaller than those from other two regimes.

\subsection{Heterogeneous models}\label{sec:heterogeneous-model}

In the following two examples we compare output from our code with the
semi-analytic formulae given in \cite{diaz08:_ana3d} using the
associated Fortran code ``Gar6more3D''. We consider domains which are
split into two layers through the $(x,y)$ plane.  The upper layer has one
set of physical properties and the lower layer another.

First we make some remarks on the semi-analytic formulae. Diaz and
Ezzani derive their formula for poroelastic wavefields in the case
that $x>0$ and $y=0$, and casually remark that the general case
follows by rotational symmetry about the $z$ axis (Equations (17)-(19)
in their paper). In particular, this symmetry would imply that the $x$
component of velocity along the $x$ axis is the same as the $y$
component of velocity along the $y$ axis which is not always the case.
However, elastic wavefields that are generated from moment tensors do
not possess this simple symmetry.  The diagonal components of the
moment tensor represent dipoles \cite{kennett01} and this is evident
in wavefield visualisations with the `split corona' phenomenon, see
Figure \ref{fig:for_adj_field_snapshots}. Poroelastic waves have even
less symmetry due to the complex coupling between solid and
fluid. This was first observed by Biot in his classic paper
\cite{biot56a} where he showed that the amplitudes of the slow P-wave
velocities for the solid and fluid components are out of phase (have
opposite sign).  Furthermore, the implemented code does not reliably
produce a solution for $z<0$ for a layered poroelastic model but,
instead, produces many untrapped LAPACK errors.  For this reason, in
section \ref{sec:poro-poro} we choose receiver locations in the upper
half domain only. For elastic-elastic coupling the implementation
works for all $z$. The example in section \ref{sec:ela-ela} therefore
contains receiver locations in the upper and lower domains. We note,
however, that the derivations for the elastic case are not documented
but presumably follow \emph{mutatis mutandis} from the poroelatic
case.

\subsubsection{Poroelastic-poroelastic}\label{sec:poro-poro}

In this experiment, the computational domain is a cube
$\Omega=[-300,300]\times[-300,300]\times[-300,300]$\,m, with the
plane $z=0$ forming the interface between two poroelastic
subdomains. Material details and derived wave speeds are given in
Table \ref{tab:params}.

We introduce a seismic source using a seismic moment tensor $M$
\cite{aki09}
\begin{displaymath}
M = 
  \begin{pmatrix}
    M_{xx} &M_{xy} & M_{xz}\\
    M_{xy} &M_{yy} & M_{yz}\\
    M_{xz} &M_{yz} & M_{zz}\\
  \end{pmatrix}
\end{displaymath}
at a point source location
\begin{equation}
  \label{eq:seismicmom}
  \g_s=\left(g_x,g_y,g_z\right)\tp=-M\cdot\nabla\delta(x_s,y_s,z_s)g(t),
\end{equation}
where $\delta$ is a Dirac delta function and $g$ is a time-dependent
source function. The source function is a Ricker wavelet with peak
frequency $f_0=20$ Hz and time delay $t_0=1.2/f_0$ and is located at
the point $(x_s,\ y_s,\ z_s)=(0,0,150)$ m. In addition, we set the
off-diagonal components of the seismic tensor $M_{xy}=M_{xz}=M_{yz}=0$
and the diagonal terms $M_{xx}=M_{yy}=M_{zz}= 10^{10}$ N$\cdot$m. The
volume source term $\g_V$ is then introduced to the model
(\ref{hyp_sys_mat}) by setting
\begin{equation}
  \label{eq:volumesource}
  \g_V=(\mathbf{0}_7,g_x,g_y,g_z,g_x,g_y,g_z)\tp
\end{equation}
where $\mathbf{0}_7$ is a $1 \times 7$ zero vector. Finally, absorbing
boundary conditions are applied across the whole boundary.

\begin{table}%
  \centering
  \caption{Material parameters and derived wave speeds used with the
    poroelastic-poroelastic case in Section
    \ref{sec:poro-poro}. }\label{tab:params}
  \begin{tabular}{c|ccc}
    \hline
    \textbf{variable name} & \textbf{symbol}  & \textbf{upper} & \textbf{lower}\\
    \hline
    solid density &  $\rhos$ (kg/m$^{3}$) & 4080 & 2700\\
    fluid density &  $\rhof$ (kg/m$^{3}$) & 1200 & 600\\
    fluid bulk modulus &  $\kappaf$ (GPa) &  5.25 & 2.0\\
    frame bulk modulus &  $\kappafr$ (GPa)&  2.0&  6.1 \\
    solid bulk modulus &  $\kappas$ (GPa) & 20.0 & 40.0 \\
    frame shear modulus &  $\mufr$ (GPa) & 6.4 & 8.0   \\
    tortuosity &  $\tau$                                &  2.0 & 2.5 \\
    porosity &  $\phi$                               &0.4 & 0.2 \\
    viscosity & $\eta$ (Pa$\cdot$s) & 0& 0\\
    \hline
    fast pressure wave speed & $\cpI$ (m/s)   & 2553 & 2990\\
    slow  pressure wave speed & $\cpII$ (m/s)  & 1097 & 844 \\
    shear wave speed          & $c_s$ (m/s)   & 1452 & 1893  \\
    \hline
  \end{tabular}
\end{table}

The computational domain $\Omega$ is partitioned by an irregular
tetrahedral grid consisting of 148187 elements and 26778 vertices
($h_{\min}=9.6$\,m and $h_{\max}=49.7$\,m). For the grid, the element
size is chosen to be 2 elements per shortest wavelength in both
subdomains. We set the basis order to 6.

\begin{figure*}%
  \centering\includegraphics[width=0.8\textwidth]{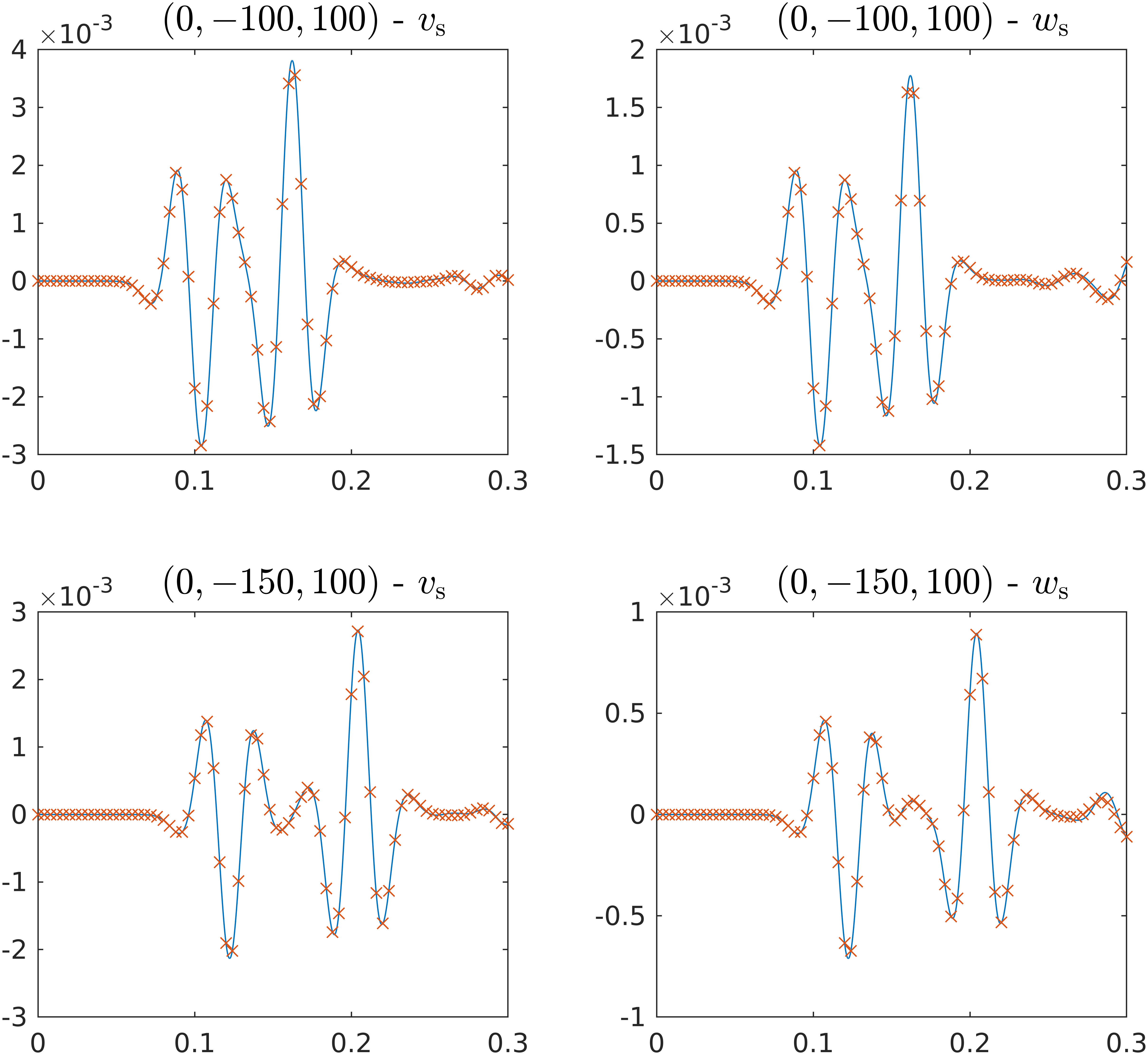}%
  \caption{Simulation of the two-layered poroelastic model of Section
    \ref{sec:poro-poro}, showing time histories of the
    velocity components $\vs$ (left), $\ws$ (right) at two receiver
    locations: DG (line) and semi-analytic (crosses). The upper row
    corresponds to receiver location $(x, y,z) = (0,-100,100)$\, m and
    the lower row to $(x, y,z) = (0,-150,100)$\,m.}
  \label{fig:2mat_signals}
\end{figure*}

The solid velocity components $\vs,\ws$ as functions of time are
shown in Figure \ref{fig:2mat_signals} at $(x,y,z) = (0,-100,100)$\,m
and $(x,y,z) = (0,-150,100)$\,m.  The signal responses show excellent
agreement with the semi-analytic solution ``Gar6more3D''
\cite{diaz08:_ana3d}. We observe the separation between the fast and
slow P-waves as the distance from the source increases. Note that the
model setup is chosen so that we do not get any unwanted reflections
from the outflow boundaries within the computed time window.
  
\subsubsection{Elastic-elastic}\label{sec:ela-ela}

The following experiment has a similar set-up to the previous one: the
computational domain is a cube
$\Omega=[-2.5,2.5]\times[-2.5,2.5]\times[-2.5,2.5]$\,m, with the plane
$z=0$ forming the interface between elastic subdomains with the same
moment tensor.  The characteristic frequency of the Ricker wavelet has
been increased to 2000 Hz and absorbing boundary conditions applied
across the whole boundary. Material densities and wave speeds are
given in Table \ref{tab:params2}.

\begin{table}%
  \centering
  \caption{Densities and  wave speeds for the
    elastic-elastic case in Section
    \ref{sec:ela-ela}. }\label{tab:params2}
  \begin{tabular}{c|ccc}
    \hline
    \textbf{variable name} & \textbf{symbol}  & \textbf{upper} & \textbf{lower}\\
    \hline
    solid density &  $\rhos$ (kg/m$^{3}$) & 2000 & 700\\
    \hline
    pressure wave speed & $c_p$ (m/s)   & 3500 & 2800\\
    shear wave speed    & $c_s$ (m/s)   & 2000 & 700  \\
    \hline
  \end{tabular}
\end{table}

The computational domain $\Omega$ is partitioned by an irregular
tetrahedral grid consisting of 35792 elements and 6858 vertices
($h_{\min}=0.12$\,m and $h_{\max}=0.95$\,m). For the grid, the element
size is chosen to be 2 elements per shortest wavelength in both
subdomains. We set the basis order to 4.

\begin{figure*}%
  \centering\includegraphics[width=0.8\textwidth]{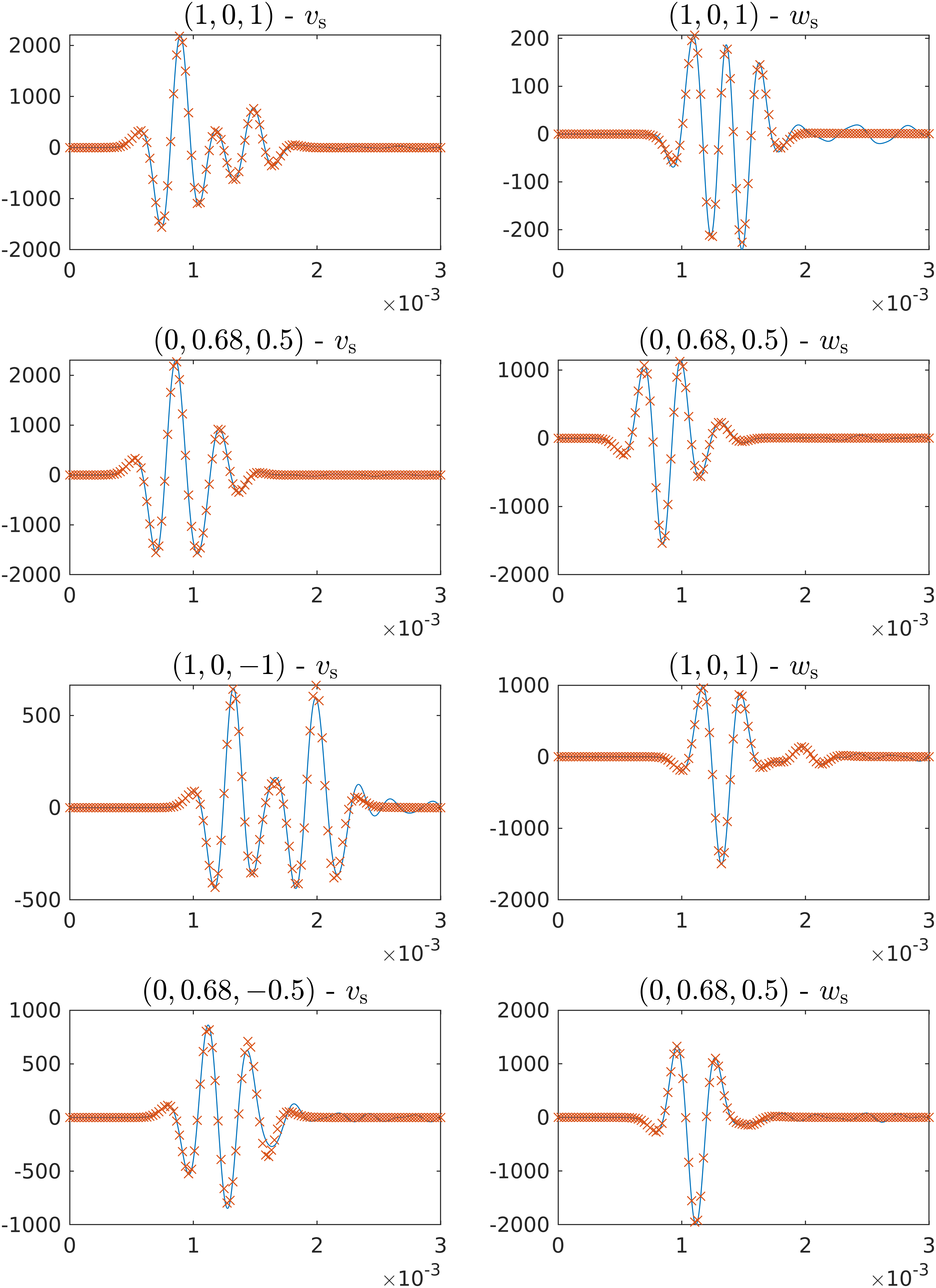}
  \caption{Simulation of the two-layered elastic model of Section
    \ref{sec:ela-ela}, showing time histories of the velocity
    components $\vs$ and $\ws$ at four locations: DG (line)
    and semi-analytic (crosses).
  }
  \label{fig:3mat_signals}
\end{figure*}

The solid velocity components $\vs$ and $\ws$ as functions of time are
shown in Figure \ref{fig:3mat_signals} for two receiver locations in
the upper layer and two in the lower half layer (coordinates: (0 1 1),
(0 0.68 0.5), (0 1 -1), (0 0.68 -0.5)). To achieve the fit we scaled
the semi-analytic output by 0.5.  The reason for this is that when
applying cylinderical symmetry to the elastic wave equation the source
term is scaled by 0.5 and this appears to be absent in the derivation
in \cite{diaz08:_ana3d}. Apart from some minor boundary reflections
for two components at the more distant locations from the source, the
signal responses show excellent agreement with the semi-analytic
solution ``Gar6more3D''.

\cleardoublepage

\subsection{Adjoint method} \label{sec:adjoint-eg}

In this section we present a simple, but illuminating, example of the
application of the adjoint method. The example was motivated by the
example for the acoustic wave equation in section 3 of
\cite{fichtner06}.  We consider an almost everywhere homogeneous
cubical domain $\Omega=[0,5]\times[0,5]\times[0,5]$\,m with one
anomalous feature in a single element containing the point
$[2.5, 2.5, 4]$\,m where the solid and fluid densities are doubled.  The
homogeneous parameters are the same as the material parameters used in
the convergence tests, Table~\ref{tab:params_conve}. A point source is
located at $[2.5, 2.5, 2]$ m and modelled as an explosive source with
$M_{xx} = M_{yy} = M_{zz} = 100$ N$\cdot$m.  The central frequency of
the Ricker wavelet is assumed to be 2000 Hz.  Velocity data is then
generated for the problem at 100 equally spaced receiver locations on
the top surface $z=5$\,m. Free surface boundary conditions were
implemented on all 6 boundary surfaces.

On the other hand the reference model is assumed to be everywhere
homogeneous with parameters given in
Table~\ref{tab:params_conve}. Letting $\boldtheta_0$ denote the
parameter space for the anomalous model, and $\boldtheta$ denote the
parameter space for the reference model, we may write the misfit
functional as

\begin{equation}
  \label{missfit_eg}
  \chi(\boldtheta) = \frac{1}{2} \sum_{i\in\mathcal{I}, r\in\mathcal{R}}\int_0^T
            [q_i(\boldtheta,x_r,t) - q_i(\boldtheta_0, x_r,t)]^2 dt
\end{equation}

\begin{figure*}%
  \centering\includegraphics[width=0.6\textwidth]{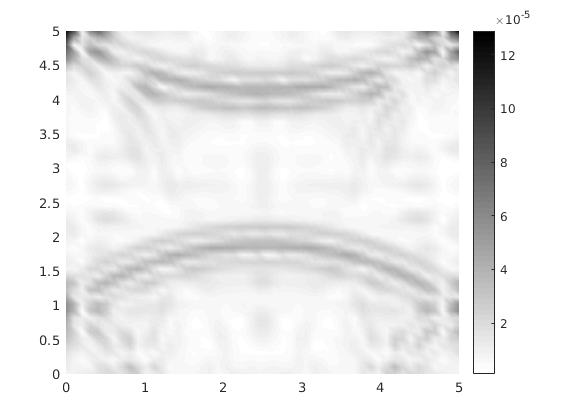}\\
  \caption{Snapshot of the forward wavefield at the final
    time $t = 3\times 10^{-3}$ seconds through the plane $x=2.5$\,m.}
  \label{fig:final_value_forward_wavefield}
\end{figure*}

\begin{figure*}%
  \centering\includegraphics[width=1\textwidth]{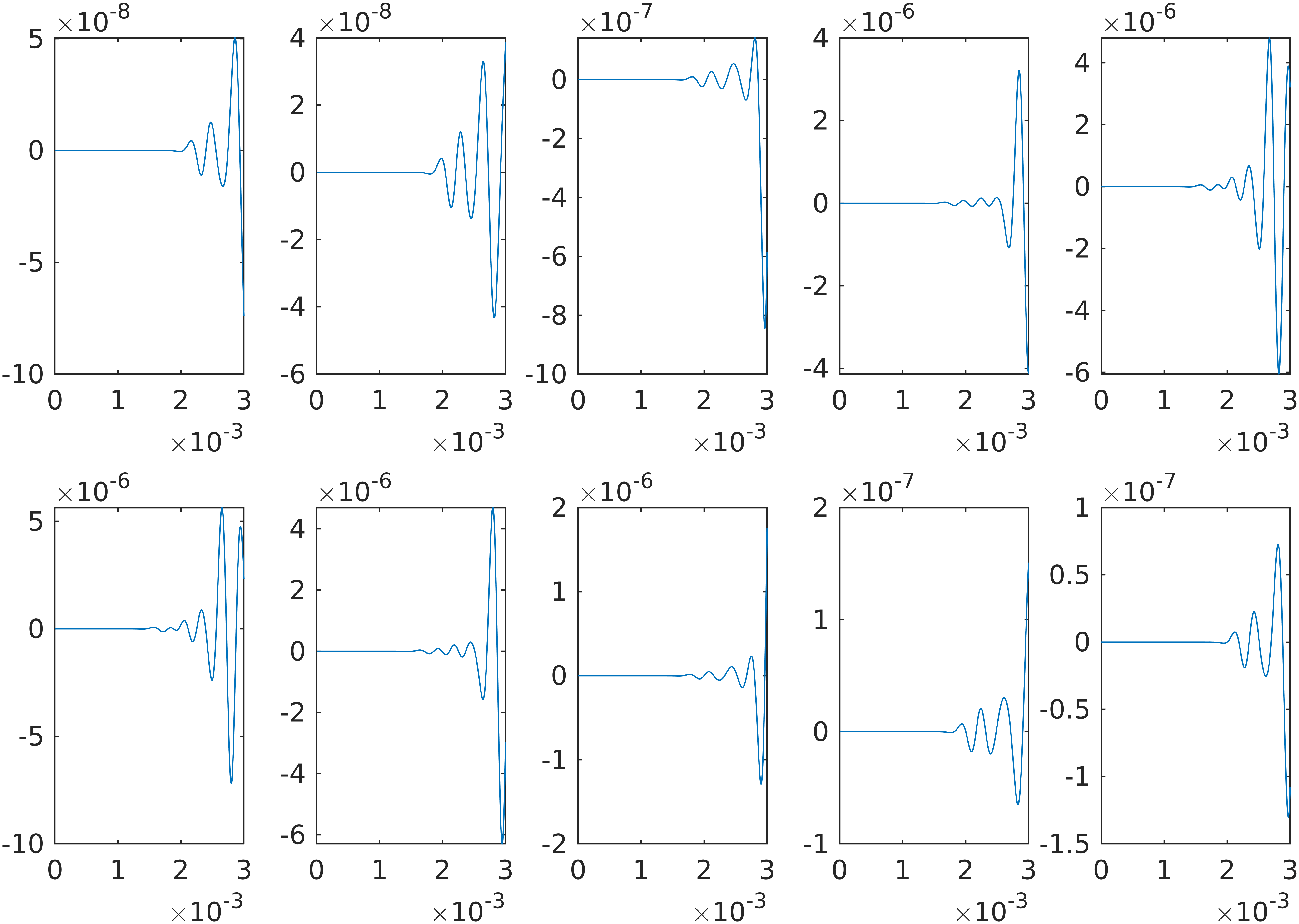}\\
  \caption{Examples of the $x$-component of adjoint source wavelets
    through the line $x=5.6$\,m.}
  \label{fig:adjoint_sources}
\end{figure*}

\begin{figure*}%
  \centering\includegraphics[width=0.45\textwidth]{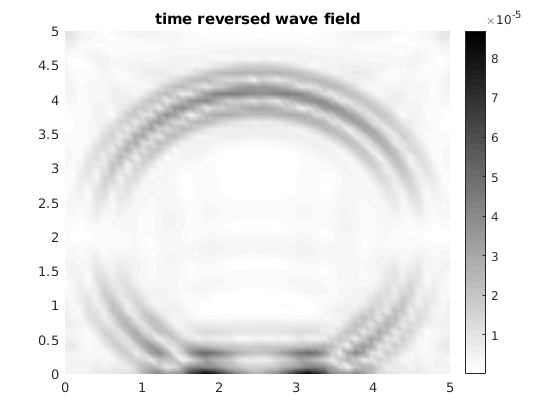}
  \includegraphics[width=0.45\textwidth]{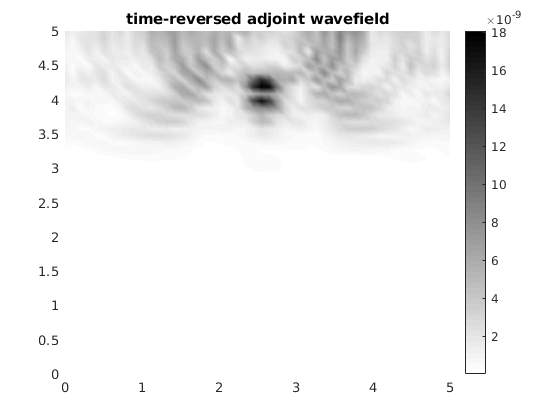}\\
  \centering\includegraphics[width=0.45\textwidth]{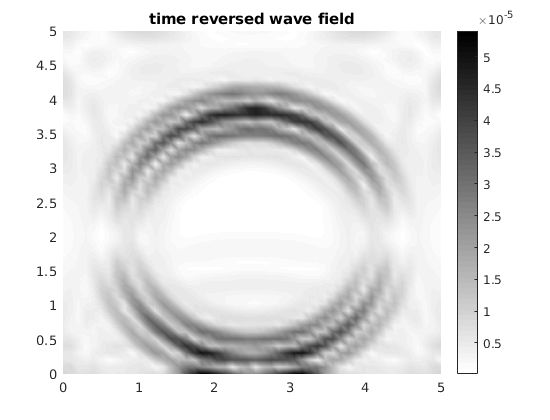}
  \includegraphics[width=0.45\textwidth]{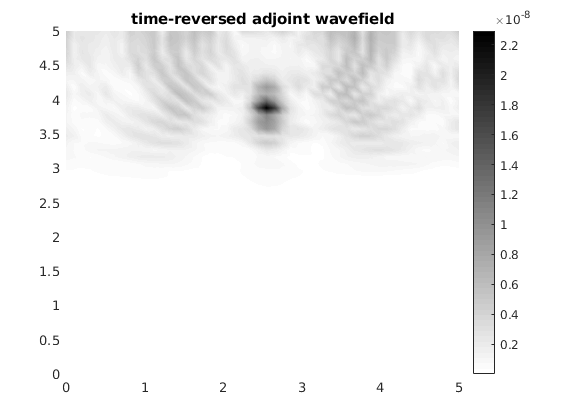}\\
  \centering\includegraphics[width=0.45\textwidth]{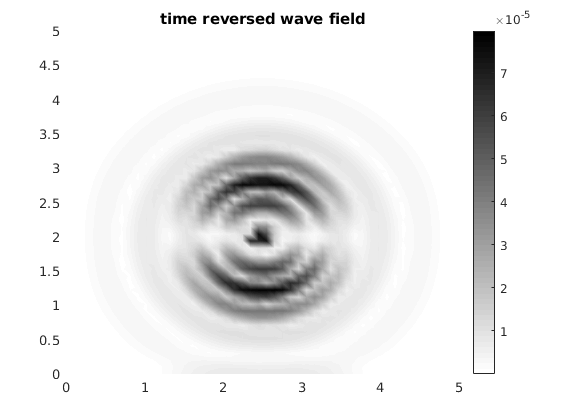}
  \includegraphics[width=0.45\textwidth]{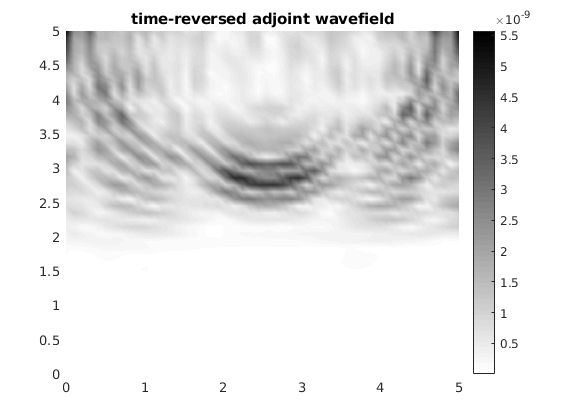}
  \caption{Snapshots of the foward (left) and adjoint (right)
    wavefields at times
    $1.85\times 10^{-3}, 1.65\times 10^{-3}, 1\times 10^{-3}$
    seconds.}
  \label{fig:for_adj_field_snapshots}
\end{figure*}

\begin{figure*}%
  \centering\includegraphics[width=0.6\textwidth]{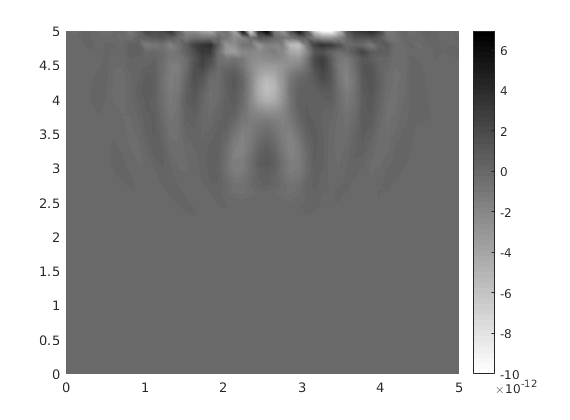}\\
  \caption{Snapshot of Fr\'echet kernel $k_{\rho_a}$ through the plane
    $x=2.5$\,m.}
  \label{fig:frechet_ker_rho_a}
\end{figure*}

The forward wavefield is propagated for $6/f_0 = 3\times 10^{-3}$
seconds and a snapshot shown in
Figure~\ref{fig:final_value_forward_wavefield}. It is evident that
considerable scattering has occurred. Examples of adjoint source
wavelets $(\q - \boldd)\delta(x-x_r)\chi_\mathcal{I}$ are shown in
Figure~\ref{fig:adjoint_sources}. Not surprisingly the central four
wavelets contain the most information since they are closest to the
anomaly, while the furthest two receivers contain further information
due to the scattering on the sides of the wave field.
Figure~\ref{fig:for_adj_field_snapshots} shown snapshots of the
forward wavefield and adjoint wavefield at times
$.85\times 10^{-3}, 1.65\times 10^{-3}, 1\times 10^{-3}$ seconds.  It
is evident that the adjoint field focuses briefly on the anomalous
feature, during which time the forward wavefront passes through the
neighbourhood.  Therefore the contribution to the Fr\'echet kernel
$k_{\rho_a}$ is greatest during this non-trivial overlapping period.
Figure~\ref{fig:frechet_ker_rho_a} shows a snapshot of $k_{\rho_a}$.
It is evident that the kernel's centre contains the anomalous element,
which extends into two tooth-like roots.  The interference near the
top surface is due to the early time overlap between the scattered
forward wavefield and the initial evolution of the adjoint wavefield.

\section{Discussion}\label{sec:discussion}

As with our previous two-dimensional work \cite{dwle17}, our principal
motivation for working in the DG framework was to obtain a forward
solver for three-dimensional poroelastic wavefields that could
accurately resolve material discontinuities. This is a necessary
feature for groundwater tomographic applications, in which abrupt
changes in porosity and permeability commonly occur between
water-bearing and non-water-bearing strata.  In our initial studies of
related inverse problems we used the well-known SPECFEM code to
simulate forward poroelastic wavefields.  However, as discussed in
section 6.2.1 of our previous paper, this approach does not naturally
resolve discontinuities in porosity, whereas the DG approach, as we
have shown in section \ref{sec: conv_anal}, naturally deals with this.
Furthermore, in applications to groundwater tomography, aquifer
permeabilities can be quite large (up to $k\sim 10^{-7}\ \text{m}^2$,
\cite{bear79}), forcing one to operate simultaneously in
high-frequency regimes (water-saturated subdomains) and low-frequency
regimes (air-saturated subdomains).
The elastic/poroelastic coupling is necessary since the usually much
slower secondary P-wave puts a very significant computational burden
on the forward solver because the mesh resolution is controlled by the
shortest wavelength.  One approach to model reduction in estimation
problems, significantly reducing the computational burden, is to make
an elastic approximation in some subdomains \cite{lahivaara14,
  lahivaara15} and, of course, the basement of an aquifer is plainly
modelled as an elastic layer.  Our implementation permits coupling
between low frequency poroelastic, high frequency poroelastic and
elastic subdomains.

With a certain loss of elegance, it is a simple extension to deal with
non-isotropic domains and to add further attenuation mechanisms for
modelling a viscoporoelastic system.  However, since poroelastic
inverse problems are extremely challenging, and we have been unable to
find a satisfactory approach to solving even modest scale problems in
two dimensions, our view is that there is still significant work to do
before tackling inverse problems for non-isotropic domains.

The adjoint method is a necessary approach to reducing the
computational burden of non-trivial inverse problems, especially those
using gradient-based approaches to minimising a misfit functional or
maximum a posteriori estimation (MAP) in the Bayesian framework.
Again the equations have been derived with some generality permitting
coupling between low and high frequency poroelastic and elastic
domains.  To our knowledge, the application of the adjoint method to
poroelastic inverse problems has been little explored
\cite{morency09}. We prefer to work with bulk parameters like the Biot
coefficient $\alpha$ and the coupling coefficient $M$ in estimation
problems since it is less cumbersome, and then use sampling to
estimate the real physical parameters of interest like porosity.  In
\cite{morency09}, on the other hand, Morency and Tromp have explored
the adjoint method for two-dimensional poroelastic problems using the
spectral element framework, and derived lengthy expressions for the
Fr\'echet kernels for the underlying physical parameters. While they
draw some parallels with the elastic case, there is much work to be
done to fully explore the utility of the adjoint method for
poroelastic inverse problems.

In our numerical simulations, for smaller examples, we used the
well-established Matlab code of Hesthaven and Warburton
\cite{hesthaven_warburton_book}. As they acknowledge in their
introduction, this becomes impractical for larger meshes; for these,
we implemented the DG algorithm in C, using MPI for parallelism and
METIS~\cite{Karypis1998} to partition the mesh. Running on a standard
desktop computer with four or six cores, this is typically faster than
Matlab by a factor of about 4 or 5. The limiting factor seems likely
to be the memory speed: in any language, the code must repeatedly
traverse arrays much larger than the system's memory caches
(Cavaglieri and Bewley \cite{CavaglieriBewley2015}, whose low-storage
IMEX schemes we used for stiff cases, mention this point in their
abstract). More important than the speed is the scalability of the MPI
code: for a large mesh, both the computational and the memory
requirements can be distributed across many nodes of a cluster. At
this point, communication costs become significant, or even dominant:
the parallel processes need to synchronise by exchanging data at every
Runge-Kutta stage (so, five times per time-step for the LSERK method
that we used for non-stiff problems) and no computation takes place
until all communication has finished. This tension between
computational and communication costs leads to a not easily predicted
optimal number of processes for any given problem. For example, on a
mesh with about 150,000 elements and 84 nodes per element (polynomial
degree 6), experimentation on the Viking cluster at the University of
York suggested that execution time would be minimised by using
somewhere around 50-60 cores; in the ever-changing environment of a
shared cluster, more precise statements are impossible.

As permeability becomes smaller, the onset of stiffness in the
low-frequency dissipative terms begins to demand unfeasibly small step
lengths in any explicit Runge-Kutta method. In these cases, we used a
hybrid implicit-explicit (IMEX) scheme in which the stiff terms are
handled by the implicit part of the scheme and the rest of the system
is handled by the explicit part.  Our formulation is ideally suited to
this type of scheme, because the equations in the implicit part can
solved simply and explicitly, entirely eliminating the extra costs
usually associated with implicit schemes. This gives convergence of
the scheme but, unlike in all other regimes, we did not observe the
convergence rates expected for the main hyperbolic system. Our
interpretation of this is that the numerical errors associated with
the dissipative terms dominate those associated with the hyperbolic
system.

\section{Conclusions}\label{sec:conclusions}

In this paper we developed a DG solver for a coupled three-dimensional
poroelastic/elastic isotropic model incorporating Biot's low- and
high-frequency regimes in Hesthaven and Warburton's framework
\cite{hesthaven_warburton_book}.  Time integration was carried out
using both low-storage explicit and (for the stiff case)
implicit-explicit Runge-Kutta schemes.  We considered free surface and
absorbing boundary conditions, where the latter were modelled as
outflows.  Numerical experiments showed that, except for very stiff
cases, the solver satisfied theoretical convergence rates. In stiff
examples, IMEX time integration gave weaker convergence rates. We
observed that the exact Riemann-problem-based numerical flux
implementation resolves naturally all material discontinuities.  We
showed that the adjoint wavefield has a natural physical
interpretation as a velocity/strain formulation of the Biot equation;
this will be further explored in a forthcoming paper.

A MATLAB implementation of the DG schemes derived in this paper to
accompany the Hesthaven-Warburton DG library is available from github:

\url{https://github.com/nickdudleyward/dg_biot_3d_1.0}

%

%

%
%
%
%

%

\end{document}